\documentclass[pra,aps,showpacs,twocolumn,longbibliography]{revtex4}
\UseRawInputEncoding

\usepackage{amssymb}
\usepackage{amsmath}
\usepackage{mathtools}
\usepackage[dvipdfm]{hyperref}
\usepackage{hyperref}
\usepackage{stmaryrd}
\usepackage{float}

\allowdisplaybreaks
\begin{document}

\title{Imaginarity of Gaussian states}
\author{Jianwei Xu}
\email{xxujianwei@nwafu.edu.cn}
\affiliation{College of Science, Northwest A$\&$F University, Yangling, Shaanxi 712100,
China}

\begin{abstract}
It has been a long-standing debate that why quantum mechanics uses complex
numbers but not only real numbers. To address this topic, in recent years,
the imaginarity theory has been developed in the way of quantum resource
theory. However, the existing imaginarity theory mainly focuses on the
quantum systems with finite dimensions. Gaussian states are widely used in
many fields of quantum physics, but they are in the quantum systems with
infinite dimensions. In this paper we establish a resource theory of
imaginarity for bosonic Gaussian states. To do so, under the Fock basis, we
determine the real Gaussian states and real Gaussian channels in terms of
the means and covariance matrices of Gaussian states. Also, we provide two
imaginary measures for Gaussian states based on the fidelity.
\end{abstract}
\maketitle
\date{\today }

\section{Introduction}
Complex numbers are widely used in both physics and mathematics. It is a
long-standing debate since the inception of quantum mechanics that why
quantum mechanics uses complex numbers but not only real numbers. To improve
this topic, recently, the imaginarity theory has been developed \cite{JPA-Gour-2018,PRL-Guo-2021,PRA-Guo-2021,Nature-2021-Acin,PRR-2021-Zhu,
PRAP-2021-Guo,QIP-2021-Li,arxiv-2022-Streltsov,PRL-2022-Li,PRA-2022-Luo,arXiv-2023-Guo}. We
consider a quantum system associated by the complex Hilbert space $H,$ and
choose an orthonormal basis $\{|j\rangle \}_{j=1}^{d}$ of $H$ with $d$ the
dimension of $H.$ Imaginarity theory is basis dependent, when we talk about
imaginarity theory, we always preset an orthonormal basis. A quantum state represented by a density operator $%
\rho $ is called real with respect to $\{|j\rangle \}_{j=1}^{d}$ if $\rho
_{jk}=\langle j|\rho |k\rangle \in R$ for all $j,k,$ here $R$ denotes the
set of all real numbers. A quantum operation \cite{Nielsen-2010-quantum} $\phi $ on $H$ is
often represented by a set of Kraus operators $\phi =\{K_{\mu }\}_{\mu }$
satisfying $\sum_{\mu }K_{\mu }^{\dagger }K_{\mu }\preceq I$, here $K_{\mu
}^{\dagger }$ is the adjoint of $K_{\mu },$ $I$ is the identity operator, $%
\sum_{\mu }K_{\mu }^{\dagger }K_{\mu }\preceq I$ means that $I-\sum_{\mu
}K_{\mu }^{\dagger }K_{\mu }\succeq 0,$ i.e., $I-\sum_{\mu }K_{\mu
}^{\dagger }K_{\mu }$ is positive semidefinite. A quantum operation $\phi
=\{K_{\mu }\}_{\mu }$ is called a quantum channel if $\sum_{\mu }K_{\mu
}^{\dagger }K_{\mu }=I.$ In imaginarity theory, an operation $\phi $ is
called real if $\phi $ can be expressed by a set of Kraus operators $\phi
=\{K_{\mu }\}_{\mu }$ and $K_{\mu }\rho K_{\mu }^{\dagger }$ is real for any
$\mu $ and any real state $\rho .$

Imaginarity theory can be viewed as a quantum resource theory. Quantum
resource theories provide a powerful way to characterize certain quantum
properties of a quantum systems \cite{IJMPB-Horodechi-2013,RMP-Gour-2019}. The well known quantum resource
theories are entanglement theory \cite{PRL-1997-Vedral,RMP-2009-Horodecki} and coherence theory \cite{PRL-Plenio-2014,RMP-Plenio-2017,PRL-2019-Brub,AQT-2021-Wu}.
Besides, other quantum resources have been developed, such as quantum thermodynamics \cite{Goold-2016,lostaglio2019introductory}, purity \cite{PRA-2003-Horodecki,PR-2015-Gour,NJP-2018-Streltsov},  nonlocality \cite{JPA-2014-De} and continuous-variable quantum resource theories \cite{NJP-2021-Brub,PRL-Takagi-2021}. A
quantum resource theory for quantum states has two basic ingredients, free
states and free operations. Resource measure and state transformation are
two main topics in a quantum resource theory for quantum states. Imaginarity
theory characterizes the property that a quantum state may be complex but
not real. In imaginarity theory, the free states are real states and free
operations are real operations. State transformations under real operations
have been extensively studied \cite{arxiv-2022-Streltsov}. Several imaginarity measures have
been proposed \cite{JPA-Gour-2018,PRL-Guo-2021,PRA-Guo-2021,QIP-2021-Li,arxiv-2022-Streltsov}. Some results of imaginary theory were experimentally
testified \cite{PRL-Guo-2021,Nature-2021-Acin,PRAP-2021-Guo,PRR-2021-Zhu,PRL-2022-Li}.

Imaginarity theory above mainly focuses on finite-dimensional quantum
states. When we attempt to apply the concepts and results of imaginarity
theory to infinite-dimensional quantum states, two problems occur. Firstly,
for the quantum states and quantum operations on infinite-dimensional
systems, there may be some ``divergence" difficulties, such as the energy of
a quantum state, then some definitions for finite-dimensional states can no
longer be well defined for infinite-dimensional states. Secondly, even if a
definition or result is still well defined for infinite-dimensional states
in a sense, there still may be a problem that this definition or result is hard to
evaluate. These problems are similar to the cases of coherence theory. In
coherence theory, the $l_{1}$ norm of coherence $C_{l_{1}}(\rho
)=\sum_{j\neq k}|\langle j|\rho |k\rangle |$ is a valid coherence measure  \cite{PRL-Plenio-2014}
and can be easily calculated for finite-dimensional states. But $C_{l_{1}}(\rho )$ may diverge for some infinite-dimensional states \cite{PRA-Fan-2016}.
In coherence theory, the relative entropy of coherence $C_{\text{r}}(\rho
)=S(\rho _{\text{diag}})-S(\rho )$ is a valid coherence measure  \cite{PRL-Plenio-2014} which can be
easily calculated for finite-dimensional states, but $C_{\text{r}}(\rho )$
is hard to calculate for some infinite-dimensional states \cite{PRA-Fan-2016,PRA-Xu-2016,PRA-2017-Paris}. Where $%
S(\rho )=-$tr($\rho \log _{2}\rho )$ is the Von Neumann entropy and $\rho _{\text{diag}}$ is the diagonal part of $\rho$.

Bosonic Gaussian states are a class of infinite-dimensional states, which
are widely used in quantum optics and quantum information theory \cite{RMP-2005-Braunstein,PR-2007-Wang,arXiv-2005-Ferraro,EPJ-2012-Olivares,
RMP-2012-Weedbrook,OSID-2014-Adesso,book-2017-Serafini}. A
Gaussian state $\rho $ is completely and conventionally described by its
mean $\overline{X}$ and covariance matrix $V$, then we write $\rho $ as $%
\rho (\overline{X},V).$ The Fock basis is the orthonormal basis spanning the complex
Hilbert space where the Gaussian states are in, then it is natural to choose
Fock basis as the fixed basis for imaginarity theory of Gaussian states. So
far, several imaginarity measures for finite-dimensional states have been
proposed, such as $I_{\text{tr}}(\rho )$ based on the trace norm \cite{JPA-Gour-2018,PRL-Guo-2021}, $%
I_{\text{r}}(\rho )$ based on the Von Neumann entropy \cite{QIP-2021-Li}, and $I_{%
\text{f}}(\rho )$ based on the fidelity \cite{PRA-Guo-2021,arxiv-2022-Streltsov}, they are defined as
\begin{eqnarray}
I_{\text{tr}}(\rho ) &=&||\rho -\rho ^{\ast }||_{\text{tr}}, \label{eq1-1} \\
I_{\text{r}}(\rho ) &=&S(\text{Re}\rho )-S(\rho ),  \label{eq1-2} \\
I_{\text{f}}(\rho ) &=&1-F(\rho ,\rho ^{\ast }),  \label{eq1-3}
\end{eqnarray}%
where $\rho ^{\ast }$ is the conjugate of $\rho ,$ $||\cdot ||_{\text{tr}}$
denotes the trace norm, $\text{Re}\rho$ is the real part of $\rho,$ $F(\rho ,\sigma )=$tr$\sqrt{\sqrt{\rho }\sigma \sqrt{\rho }}$ is the fidelity of states $\rho $ and $\sigma$ \cite{RMP-1976-Uhlmann,JMP-1994-Jozsa}. We consider
whether Eqs. (\ref{eq1-1},\ref{eq1-2},\ref{eq1-3}) are applicable to Gaussian states. Till now, to calculate $\rho ^{\ast },$ Re$\rho $ and $||\rho -\rho ^{\ast }||_{\text{tr}}$ for general Gaussian states is very
hard since it is hard to express general Gaussian states in Fock basis
\cite{PRA-Fan-2016,PRA-Xu-2016,PRA-2017-Paris,PRA-2019-Quesada}. $F(\rho ,\sigma )$ has a closed expression for Gaussian states $%
\rho $ and $\sigma $ in terms of their means and covariance matrices \cite{PRL-2015-Banchi},
but we do not know whether $\rho ^{\ast }$ is a Gaussian state. Moreover, we
do not even know which Gaussian states are real in terms of means and
covariances.

In this paper we study the imaginarity of bosonic Gaussian states. We will
establish a resource theory of imaginarity for bosonic Gaussian states.
This paper is structured as follows. In section 2, we determine the
conditions for real Gaussian states and the conjugate of a Gaussian state
under Fock basis in terms of means and covariances. In section 3, we
characterize the structure of real Gaussian channels. In section 4, we
provide two imaginary measures for Gaussian states based the fidelity, they
all have closed expressions. Section 5 is a brief summary and outlook. For
structural clarity, we focus on stating the theoretical framework and results in
main text, and put most of the proofs to the Appendices.

\section{Real Gaussian states and the conjugate of a Gaussian state}

In this section we determine the real Gaussian states and the conjugate of a
Gaussian state. We first recall some basics and give the notation we will
use for Gaussian states. We denote the one-mode Fock basis by $\{|j\rangle
\}_{j=0}^{\infty }$ with $j\in \{0,1,2,3,...\},$ $\{|j\rangle
\}_{j=0}^{\infty }$ is an orthonormal basis spanning the complex Hilbert
space $\overline{H}.$ $\overline{H}$ is a countable but infinite-dimensional
complex Hilbert space. The $N$-mode Fock basis is $\{|j\rangle
\}_{j}^{\otimes m},$ the $N$-fold tensor product of $\{|j\rangle
\}_{j=0}^{\infty },$ and $\{|j\rangle \}_{j}^{\otimes m}$ spans the complex
Hilbert space $\overline{H}^{\otimes N}=\otimes _{l=1}^{N}\overline{H}_{l}$
with each $\overline{H}_{l}=\overline{H}.$ On each $\overline{H}_{l},$ the
bosonic field operators: annihilation operator $\widehat{a}_{l}$ and the
creation operator $\widehat{a}_{l}^{\dagger }$ are defined as
\begin{eqnarray}
\widehat{a}_{l}|0\rangle &=&0, \ \  \widehat{a}_{l}|j\rangle =\sqrt{j}|j-1\rangle
\text{ for }j\geq 1; \label{eq2-1} \\
\widehat{a}_{l}^{\dagger }|j\rangle &=&\sqrt{j+1}|j+1\rangle \text{ for }%
j\geq 0. \label{eq2-2}
\end{eqnarray}%
We arrange $\{\widehat{a}_{l},\widehat{a}_{l}^{\dagger }\}_{l=1}^{N}$ as a
vector as
\begin{eqnarray}
\widehat{A} &=&(\widehat{a}_{1},\widehat{a}_{1}^{\dagger },\widehat{a}_{2},%
\widehat{a}_{2}^{\dagger },...,\widehat{a}_{N},\widehat{a}_{N}^{\dagger
})^{T} \notag \\
&=&(\widehat{A}_{1},\widehat{A}_{2},\widehat{A}_{3},\widehat{A}_{4},...,%
\widehat{A}_{2N-1},\widehat{A}_{2N})^{T}, \label{eq2-3}
\end{eqnarray}%
with $T$ standing for the transposition.

From bosonic field operators $\{\widehat{a}_{l},\widehat{a}_{l}^{\dagger
}\}_{l=1}^{N}$ we can define the quadrature field operators $\{\widehat{q}%
_{l},\widehat{p}_{l}^{\dagger }\}_{l=1}^{N}$ as
\begin{equation}
\widehat{q}_{l}=\widehat{a}_{l}+\widehat{a}_{l}^{\dagger }, \ \
\widehat{p}_{l}=-i(\widehat{a}_{l}-\widehat{a}_{l}^{\dagger }),  \label{eq2-4}
\end{equation}%
where $i=\sqrt{-1}.$ We arrange $\{\widehat{q}_{l},\widehat{p}_{l}^{\dagger }\}_{l=1}^{N}$ as a
vector as
\begin{eqnarray}
\widehat{X} &=&(\widehat{q}_{1},\widehat{p}_{1},\widehat{q}_{2},\widehat{p}%
_{2},...,\widehat{q}_{N},\widehat{p}_{N})^{T} \notag  \\
&=&(\widehat{X}_{1},\widehat{X}_{2},\widehat{X}_{3},\widehat{X}_{4},...,%
\widehat{X}_{2N-1},\widehat{X}_{2N})^{T}.  \label{eq2-5}
\end{eqnarray}

Under these definitions, we obtain the canonical commutation relations that
\begin{eqnarray}
\lbrack \widehat{A}_{l},\widehat{A}_{m}] &=&\Omega _{lm}, \label{eq2-6} \\
\lbrack \widehat{X}_{l},\widehat{X}_{m}] &=&2i\Omega _{lm}, \label{eq2-7}
\end{eqnarray}
where $[\widehat{A}_{l},\widehat{A}_{m}]=\widehat{A}_{l}\widehat{A}_{m}-%
\widehat{A}_{m}\widehat{A}_{l}$ is the commutator of $\widehat{A}_{l}$ and $%
\widehat{A}_{m},$ $\Omega _{lm}$ is the element of the $2N\times 2N$ matrix $%
\Omega $ with
\begin{eqnarray}
\Omega=\oplus _{n=1}^{N}\omega ,  \ \
\omega=\left(
\begin{array}{cc}
0 & 1 \\
-1 & 0%
\end{array}%
\right).  \label{eq2-8}
\end{eqnarray}

A quantum state $\rho $ in $\overline{H}^{\otimes N}$ can be characterized
by its characteristic function
\begin{equation}
\chi (\rho ,\xi )=\text{tr}[\rho D(\xi )],  \label{eq2-9}
\end{equation}%
where $D(\xi )$ is the displacement operator
\begin{eqnarray}
D(\xi ) &=&\exp (i\widehat{X}^{T}\Omega \xi ),  \label{eq2-10} \\
\xi  &=&(\xi _{1},\xi _{2},...,\xi _{2N})^{T}\in R^{2N}.  \label{eq2-11}
\end{eqnarray}

For state $\rho $ in $\overline{H}^{\otimes N},$ the mean of $\rho $
is
\begin{equation}
\overline{X}=\text{tr}(\rho \widehat{X})=(\overline{X}_{1},\overline{X}%
_{2},...,\overline{X}_{2N})^{T}; \label{eq2-12}
\end{equation}%
the covariance matrix $V$ is defined by its elements
\begin{equation}
V_{lm}=\frac{1}{2}\text{tr}(\rho \{\Delta \widehat{X}_{l},\Delta \widehat{X}%
_{m}\})   \label{eq2-13}
\end{equation}%
with $\Delta \widehat{X}_{l}=\widehat{X}_{l}-\overline{X}_{l}$, and $%
\{\Delta \widehat{X}_{l},\Delta \widehat{X}_{m}\}=\Delta \widehat{X}%
_{l}\Delta \widehat{X}_{m}+\Delta \widehat{X}_{m}\Delta \widehat{X}_{l}$ is
the anticommutator of $\Delta \widehat{X}_{l}$ and $\Delta \widehat{X}_{m}.$
The covariance matrix $V=V^{T}$ is a $2N\times 2N$ real and symmetric matrix
which must satisfy the uncertainty principle \cite{PRA-1994-Simon}
\begin{equation}
V+i\Omega \succeq 0.   \label{eq2-14}
\end{equation}%
Note that $V+i\Omega \succeq 0$ implies $V\succ 0$
meaning that $V$ is positive definite.

With these preparations, we turn to the definition of Gaussian states. A
quantum state $\rho $ in $\overline{H}^{\otimes N}$ is called an $N$-mode
Gaussian state if its characteristic function has the Gaussian form
\begin{equation}
\chi (\rho ,\xi )=\exp [-\frac{1}{2}\xi ^{T}(\Omega V\Omega ^{T})\xi
-i(\Omega \overline{X})^{T}\xi ], \label{eq2-15}
\end{equation}%
where $\overline{X}$ is the mean of $\rho $ and $V$ is the covariance matrix
of $\rho .$ The Gaussian state $\rho $ is determined by its characteristic
function $\chi (\rho ,\xi )$ via the inverse relation \cite{book-2017-Serafini}
\begin{equation}
\rho =\int \frac{d^{2N}\xi }{\pi ^{N}}\chi (\rho ,\xi )D(-\xi ),  \label{eq2-16}
\end{equation}%
where $\int =\int_{-\infty }^{\infty }.$ $\overline{X}$ and $V$ with Eq. (\ref{eq2-15})
completely determine the Gaussian state $\rho $ \cite{PRA-1994-Simon}, thus we write $%
\rho $ as $\rho (\overline{X},V).$

Now we consider the question that under what conditions on $\overline{X}$
and $V,$ $\rho $ is a real Gaussian state, i.e., $\langle j_{1}|\langle
j_{2}|...|j_{N}|\rho |k_{1}\rangle |k_{2}\rangle ...|k_{N}\rangle \in R$ for
any Fock basis vectors $\{|j_{1}\rangle ,|j_{2}\rangle |,...,|j_{N}\rangle
;|k_{1}\rangle ,|k_{2}\rangle ,...,|k_{N}\rangle \}.$ For this question, we
have Theorem 1 below, we provide a proof for Theorem 1 in Appendix A.

\emph{Theorem 1.} The $N$-mode Gaussian state $\rho (\overline{X},V)$ is
real if and only if
\begin{eqnarray}
\overline{X}_{2l} &=&0\text{ for }l\in \{1,2,...,N\},  \label{eq2-17} \\
V_{2l-1,2m} &=&0\text{ for }l,m\in \{1,2,...,N\}. \label{eq2-18}
\end{eqnarray}

If one of $\{\overline{X}_{2l},V_{2l-1,2m}\}_{l,m=1}^{N}$ is nonzero, then
there exists $\langle j_{1}|\langle j_{2}|...|j_{N}|\rho |k_{1}\rangle
|k_{2}\rangle ...|k_{N}\rangle \notin R$ for $\rho (\overline{X},V)$, $\rho (%
\overline{X},V)$ is called not real. When $\rho (\overline{X},V)$ is not
real, we further ask how about the conjugate $\rho ^{\ast }$ of $\rho (%
\overline{X},V).$ Is $\rho ^{\ast }$ still Gaussian state? If $\rho ^{\ast }$
is still a Gaussian state, then how about the mean and covariance matrix of $%
\rho ^{\ast }.$ Theorem 2 below will answer these questions, we provide a
proof for Theorem 2 in Appendix B.

\emph{Theorem 2.} For $N$-mode Gaussian state $\rho (\overline{X},V),$ the
conjugate state of $\rho (\overline{X},V)$ is still a Gaussian state. We
denote the conjugate state of $\rho (V,\overline{X})$ by $\rho ^{\ast }(%
\overline{X}^{\prime },V^{\prime })$ with the mean $\overline{X}^{\prime }$
and covariance matrix $V^{\prime },$ then
\begin{eqnarray}
\overline{X}_{l}^{\prime } &=&(-1)^{l+1}\overline{X}_{l},l\in \{1,2,...,2N\},  \label{eq2-19}
\\
V_{lm}^{\prime } &=&(-1)^{l+m}V_{lm},l,m\in \{1,2,...,2N\}.  \label{eq2-20}
\end{eqnarray}

With Theorem 1, Theorem 2, and Eqs. (\ref{eq2-5},\ref{eq2-13}), we see that, for $N$-mode Gaussian state $\rho
(\overline{X},V),$ the real part of $\rho ,$
$\text{Re}\rho =\frac{\rho +\rho ^{\ast }}{2},$
has the mean $\frac{\overline{X}+\overline{X}^{\prime }}{2}$ and covariance
matrix $\frac{V+V^{\prime }}{2},$ and $(\frac{\overline{X}+\overline{X}%
^{\prime }}{2},\frac{V+V^{\prime }}{2})$ determines a real Gaussian state
since
\begin{equation}
\frac{V+V^{\prime }}{2}+i\Omega =\frac{(V+i\Omega )+(V^{\prime }+i\Omega )}{2%
}\succeq 0.  \label{eq2-21}
\end{equation}%
Then one may ask the question that
whether Re$\rho $ is a Gaussian state. The answer of this question is
negative. That is, if $\rho $ is a Gaussian state, Re$\rho $ is not a
Gaussian state in general. We can check this fact by the Glauber coherent
state in Example 2 below. With this observation, for $N$-mode Gaussian state
$\rho (\overline{X},V),$ we define a real Gaussian state $\overline{\rho }$
having the mean $\frac{\overline{X}+\overline{X}^{\prime }}{2}$ and
covariance matrix $\frac{V+V^{\prime }}{2},$ we write $\overline{\rho }$ as $%
\overline{\rho }(\frac{\overline{X}+\overline{X}^{\prime }}{2},\frac{%
V+V^{\prime }}{2}).$ We call $\overline{\rho }(\frac{\overline{X}+\overline{X%
}^{\prime }}{2},\frac{V+V^{\prime }}{2})$ the real Gaussian state induced by
the Gaussian state $\rho (\overline{X},V).$ Obviously, for Gaussian state $%
\rho (\overline{X},V),$ we have
\begin{equation}
\overline{\rho }=(\overline{\rho })^{\ast }=\overline{\rho ^{\ast }},  \label{eq2-22}
\end{equation}%
and $\rho $ is real (i.e., $\rho =$Re$%
\rho $) if and only if $\rho =\overline{\rho }.$ In general, Re$\rho \neq \overline{\rho }$ for Gaussian state $\rho (\overline{X},V).$

\section{Real Gaussian channels}

A Gaussian channel $\phi $ on $\overline{H}^{\otimes N}$ can be represented
by $\phi =(d,T,N)$, here $d=(d_{1},d_{2},...,d_{2N})^{T}\in R^{2N},$ $T$ and
$N=N^{T}$ are $2N\times 2N$ real matrices. $\phi =(d,T,N)$ maps the Gaussian state $\rho (\overline{X},V)$ to the Gaussian state with mean and covariance matrix as
\begin{eqnarray}
\overline{X}\rightarrow T\overline{X}+d, \ \ V\rightarrow TVT^{T}+N, \label{eq3-1}
\end{eqnarray}
and $\phi =(d,T,N)$ fulfils the completely positivity condition
\begin{equation}
N+i\Omega -iT\Omega T^{T}\succeq 0.  \label{eq3-2}
\end{equation}

We then define that a Gaussian channel is real if it maps any real Gaussian
state to real Gaussian state. For the structure of real Gaussian channels we
have Theorem 3 below, we provide a proof for Theorem 3 in Appendix C.

\emph{Theorem 3.} The $N$-mode Gaussian channel $\phi =(d,T,N)$ is real if
and only if
\begin{eqnarray}
d_{2l} &=&0\text{ for }l\in \{1,2,...,N\},  \label{eq3-3} \\
N_{2l-1,2m} &=&0\text{ for }l,m\in \{1,2,...,N\}, \label{eq3-4}
\end{eqnarray}
one of Eqs. (\ref{eq3-5},\ref{eq3-6}) below
\begin{eqnarray}
T_{2l,2m-1} &=&T_{2l,2m}=0\text{ for }l,m\in \{1,2,...,N\},  \label{eq3-5} \\
T_{2l-1,2m} &=&T_{2m,2l-1}=0\text{ for }l,m\in \{1,2,...,N\}, \ \  \label{eq3-6}
\end{eqnarray}
and Eq. (\ref{eq3-2}).

We discuss the properties of real Gaussian channels. If a real Gaussian
channel $\phi $ fulfils Eq. (\ref{eq3-5}), we call it a completely real Gaussian
channel. If a real Gaussian channel $\phi $ fulfils Eq. (\ref{eq3-6}), we call it a
covariant real Gaussian channel. The meanings of these definitions are
explained in Theorem 4 below. We give a proof for Theorem 4 in Appendix D.
In particular, if a real Gaussian channel $\phi $ fulfils both Eqs. (\ref{eq3-5},\ref{eq3-6}), we
call it a covariant and completely real Gaussian channel. Such a
classification of real Gaussian channels is shown in Figure 1.

\emph{Theorem 4. }If $\phi $ is a completely real Gaussian channel, then $%
\phi (\rho )$ is real for any Gaussian state $\rho .$ If $\phi $ is a
covariant real Gaussian channel, then for any Gaussian state $\rho ,$ we
have
\begin{eqnarray}
\lbrack \phi (\rho )]^{\ast } &=&\phi (\rho ^{\ast }),  \label{eq3-7} \\
\overline{\phi (\rho )} &=&\phi (\overline{\rho }).  \label{eq3-8}
\end{eqnarray}

\begin{figure}
\includegraphics[width=8cm,bb=60 70 700 400,clip]{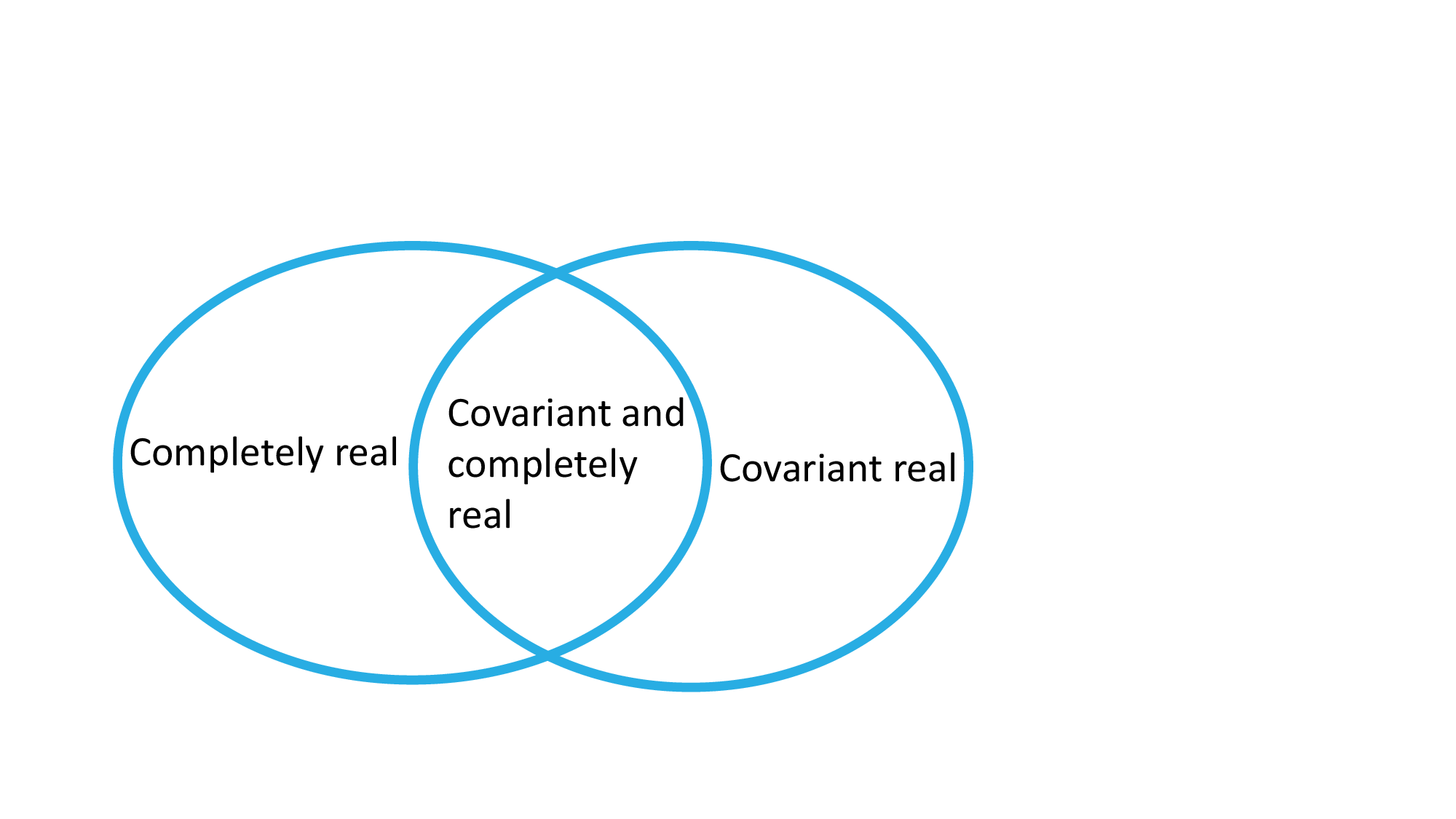}
\caption{Classification of real Gaussian channels.}
\end{figure}

\section{Imaginarity measures of Gaussian states}

An imaginarity measure $M(\rho )$ for $N$-mode Gaussian states is a real valued functional
on Gaussian states. In the spirit of quantum resource theory, we propose that any imaginarity measure $M(\rho )$ for $N$-mode Gaussian states should satisfy the following two conditions.

(M1). Faithfulness: $M(\rho )\geq 0$ for any state $\rho $ and $M(\rho )=0$
if and only if $\rho $ is real.

(M2). Monotonicity: $M(\phi (\rho ))\leq M(\rho )$ for any state $\rho $ and
any real Gaussian channel $\phi .$

We provide two imaginarity measures based on the fidelity for Gaussian
states in Theorem 5 below. We give a proof for Theorem 5
in Appendix E.

\emph{Theorem 5.} For any $N$-mode Gaussian state $\rho (\overline{X},V),$
\begin{eqnarray}
M(\rho )&=&1-F(\rho ,\rho ^{\ast }), \label{eq4-1} \\
M^{\prime }(\rho )&=&1-F(\rho ,\overline{\rho })  \label{eq4-2}
\end{eqnarray}%
are all imaginarity measures, i.e., $M(\rho )$ and $M^{\prime }(\rho )$ all satisfy (M1) and (M2).

From the definitions of $M(\rho )$ and $M^{\prime }(\rho )$, we see that $M(\rho )$ and $M^{\prime }(\rho )$ have the property of conjugation invariance
\begin{equation}
M(\rho )=M(\rho^{*} ), \ \  M^{\prime }(\rho )=M^{\prime }(\rho^{*} ).   \label{eq4-3}
\end{equation}%

It is shown that $1-F(\rho ,\rho ^{\ast })$ in Eq. (\ref{eq1-3})
is a valid imaginarity measure for finite-dimensional states  \cite{PRA-Guo-2021,arxiv-2022-Streltsov}.
We have shown that if $\rho $ is a Gaussian state, then $\rho ^{\ast }$ and $\overline{\rho}$ are all Gaussian states. Then the calculation of $M(\rho )$ and $M^{\prime }(\rho )$ is about the calculation of the fidelity for two Gaussian states. The expression of the fidelity $F(\rho ,\sigma )$ for two Gaussian states $\rho$ and $\sigma$ has been studied for many years \cite{JPA-1998-Scutaru,PRA-2003-Marian,PRA-2008-Marian,PRA-2005-Nha,PRA-2006-Olivares,PRA-2012-Marian,PRL-2015-Banchi}, and in Ref. \cite{PRL-2015-Banchi} an explicit expression of $F(\rho ,\sigma )$ for any two $N$-mode Gaussian states was provided. Consequently, $M(\rho )$ and $M^{\prime }(\rho )$ have explicit expressions via the explicit expression of $F(\rho ,\sigma )$ for any two $N$-mode Gaussian states \cite{PRL-2015-Banchi}. Below we discuss some special one-mode Gaussian states to demonstrate the calculation of $M(\rho )$ and $M^{\prime }(\rho ).$

For any two one-mode Gaussian states $\rho (\overline{X},V)$ and $\sigma (%
\overline{Y},W),$ the fidelity $F(\rho ,\sigma )$ has the expression \cite%
{PRA-2012-Marian,PRL-2015-Banchi}
\begin{eqnarray}
F(\rho ,\sigma ) &=&\frac{\exp [-\frac{1}{4}(\overline{X}-\overline{Y}%
)^{T}(V+W)^{-1}(\overline{X}-\overline{Y})]}{\sqrt{\sqrt{\det (\frac{V+W}{2}%
)+\Lambda }-\sqrt{\Lambda }}}, \ \ \ \   \label{eq4-4} \\
\Lambda &=&4\det (\frac{V+i\Omega }{2})\det (\frac{W+i\Omega }{2}).
\label{eq4-5}
\end{eqnarray}%
With these expressions we can directly calculate $M(\rho )$ and $M^{\prime }(\rho )$ for any one-mode Gaussian state $\rho$.

\emph{Corollary 1.} For one-mode Gaussian state $\rho (\overline{X},V),$ the
imaginarity measures $M(\rho )$ in Eq. (\ref{eq4-1}) and $M'(\rho )$ in Eq. (\ref{eq4-2}) become
\begin{eqnarray}
M(\rho )&=&1-\frac{\exp (-\frac{\overline{X}_{2}^{2}}{2V_{22}})}{\sqrt{\sqrt{%
V_{11}V_{22}+\Lambda}-\sqrt{\Lambda}}},   \label{eq4-6} \\
\Lambda&=&\frac{(V_{11}V_{22}-V_{12}^{2}-1)^{2}}{4};  \label{eq4-7} \\
M'(\rho )&=&1-\frac{\exp [-\frac{V_{11}\overline{X}_{2}^{2}}{2(4V_{11}V_{22}-V_{12}^{2})}]}{\sqrt{\sqrt{%
(V_{11}V_{22}-\frac{1}{4}V_{12}^{2})+\Lambda'}-\sqrt{\Lambda'}}},     \label{eq4-8}   \\
\Lambda'&=&\frac{1}{4}(V_{11}V_{22}-V_{12}^{2}-1)(V_{11}V_{22}-1).    \label{eq4-9}
\end{eqnarray}

We discuss some classes of special one-mode Gaussian states, the thermal
states, the Glauber coherent states and the squeezed states. These classes of
Gaussian states are widely used in quantum optics and quantum information
theory. For one-mode case, we also write the Fock basis $\{|j\rangle
\}_{j=0}^{\infty }=\{|n\rangle \}_{n=0}^{\infty },$ and the creation and
annihilation operator as $\widehat{a}_{1}=\widehat{a},$ $\widehat{a}%
_{1}^{\dagger }=\widehat{a}^{\dagger }.$

\emph{Example 1.} Consider the one-mode thermal state
\begin{equation}
\rho _{\text{th}}(\overline{n})=\sum_{n=0}^{\infty }\frac{\overline{n}^{n}}{(%
\overline{n}+1)^{n+1}}|n\rangle \langle n|,  \label{eq4-10}
\end{equation}%
with $\overline{n}=$tr($\widehat{a}^{\dagger }\widehat{a}\rho _{\text{th}}(%
\overline{n})$) the mean number of $\rho _{\text{th}}(\overline{n}).$ The
mean of $\rho _{\text{th}}(\overline{n})$ is $\overline{X}=(0,0)^{T},$ the
covariance matrix of $\rho _{\text{th}}(\overline{n})$ is $V=(2\overline{n}%
+1)\left(
\begin{array}{cc}
1 & 0 \\
0 & 1%
\end{array}%
\right) .$ Then Eqs. (\ref{eq4-6},\ref{eq4-7},\ref{eq4-8},\ref{eq4-9}) yield $M(\rho _{\text{th}}(\overline{n}))=M'(\rho _{\text{th}}(\overline{n}))=0.$
In fact $M(\rho _{\text{th}}(\overline{n}))=M'(\rho _{\text{th}}(\overline{n}))=0$ is an obvious result, since the
matrix elements of $\rho _{\text{th}}(\overline{n})$ are all real in Fock
basis, i.e., $\rho _{\text{th}}(\overline{n})$ is a real Gaussian state.

\emph{Example 2. }Consider the one-mode Glauber coherent state
\begin{equation}
|\alpha \rangle =D(\alpha )|0\rangle =e^{-\frac{|\alpha |^{2}}{2}%
}\sum_{n=0}^{\infty }\frac{\alpha ^{n}}{\sqrt{n!}}|n\rangle    \label{eq4-11}
\end{equation}%
with $\alpha $ any complex number. The mean of $|\alpha \rangle \langle
\alpha |$ is $\overline{X}=(2\text{Re}\alpha ,2\text{Im}\alpha )^{T},$ the
covariance matrix of $|\alpha \rangle \langle \alpha |$ is $V=\left(
\begin{array}{cc}
1 & 0 \\
0 & 1%
\end{array}%
\right).$ Then Eqs. (\ref{eq4-6},\ref{eq4-7},\ref{eq4-8},\ref{eq4-9}) yield
\begin{eqnarray}
M(|\alpha \rangle )&=&1-e^{-2(\text{Im}\alpha )^{2}};   \label{eq4-12} \\
M'(|\alpha \rangle )&=&1-e^{-\frac{(\text{Im}\alpha )^{2}}{2}}.    \label{eq4-13}
\end{eqnarray}
We see that, $M(|\alpha \rangle )>M'(|\alpha \rangle )>0$ when $\alpha \notin R,$ $M(|\alpha
\rangle )=M'(|\alpha \rangle )=0$ if and only if $\alpha \in R,$ $M(|\alpha \rangle )$ and $M(|\alpha \rangle )$ increase as $|\text{Im}\alpha |$ increases, $M(|\alpha \rangle )$ and $M'(|\alpha \rangle )$ are independent of $%
\text{Re}\alpha .$ We depict Eqs. (\ref{eq4-12},\ref{eq4-13}) in Figure 2.

\begin{figure}
\includegraphics[width=8cm]{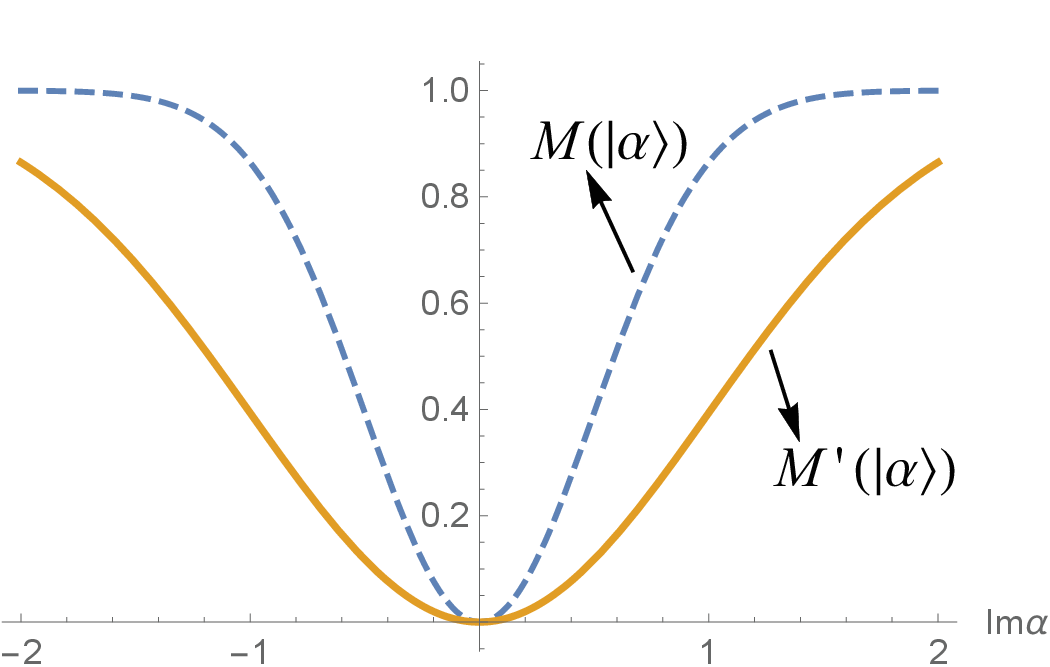}
\caption{$M(|\alpha \rangle )$ and $M'(|\alpha \rangle )$ versus $\text{Im}\alpha $ in Eqs. (\ref{eq4-12},\ref{eq4-13}).}
\end{figure}

\emph{Example 3.} Consider the one-mode squeezed state
\begin{eqnarray}
|\zeta \rangle  &=&\exp [\frac{1}{2}(\zeta ^{\ast }\widehat{a}^{2}-\zeta
\widehat{a}^{\dagger 2})]|0\rangle   \label{eq4-14} \\
&=&\frac{1}{\sqrt{\cosh |\zeta |}}\sum_{n=0}^{\infty }(-e^{i\theta }\tanh
|\zeta |)^{n}\frac{\sqrt{(2n)!}}{2^{n}n!}|2n\rangle,   \ \ \label{eq4-15}
\end{eqnarray}%
with $\zeta $ any complex number and $\zeta =|\zeta |e^{i\theta }$ its polar
form. $\exp [\frac{1}{2}(\zeta ^{\ast }\widehat{a}^{2}-\zeta
\widehat{a}^{\dagger 2})]$ is the squeezing operator. The mean of $|\zeta \rangle \langle \zeta |$ is $\overline{X}=(0,0)^{T},$
the covariance matrix $V$ of $|\zeta \rangle \langle \zeta |$ is
\begin{eqnarray}\label{eq4-16}
\begin{cases}
V_{11}=\cosh(2|\zeta |)+\cos \theta \sinh(2|\zeta |) \\
V_{12}=V_{21}=\sin \theta \sinh(2|\zeta|)  \\
V_{22}=\cosh(2|\zeta |)-\cos \theta \sinh(2|\zeta|).
\end{cases}
\end{eqnarray}
Then Eqs. (\ref{eq4-6},\ref{eq4-7},\ref{eq4-8},\ref{eq4-9}) yield
\begin{eqnarray}
&&M(|\zeta \rangle )=1-\frac{1}{\sqrt[4]{1+\sin ^{2}\theta
\sinh ^{2}(2|\zeta |)}};   \label{eq4-17} \\
&&M'(|\zeta \rangle )=1-\frac{1}{\sqrt[4]{1+\frac{3}4{}\sin ^{2}\theta
\sinh ^{2}(2|\zeta |)}}.    \label{eq4-18}
\end{eqnarray}
We see that, if $\zeta \notin R$ then $M(|\zeta \rangle )>M'(|\zeta \rangle )>0$; $M(|\zeta
\rangle )$ and $M'(|\zeta
\rangle )$ increase as $|\zeta |$ increases; $M(|\zeta \rangle )=M'(|\zeta \rangle )=0$ if and
only if $\zeta \in R.$ We depict Eq. (\ref{eq4-17}) in Figure 3, and compare Eqs. (\ref{eq4-17},\ref{eq4-18}) in Figure 4.

\begin{figure}
\includegraphics[width=8cm]{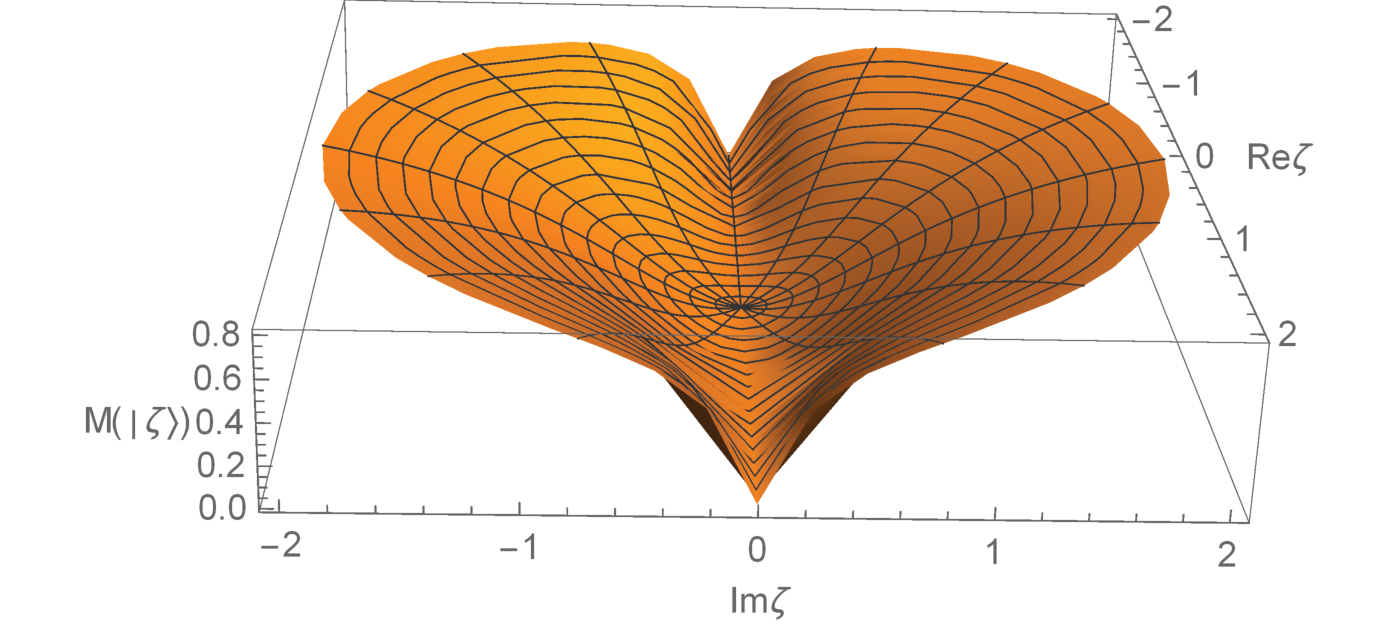}
\caption{$M(|\zeta \rangle )$ in Eq. (\ref{eq4-17}).}
\end{figure}

\begin{figure}
\includegraphics[width=8cm]{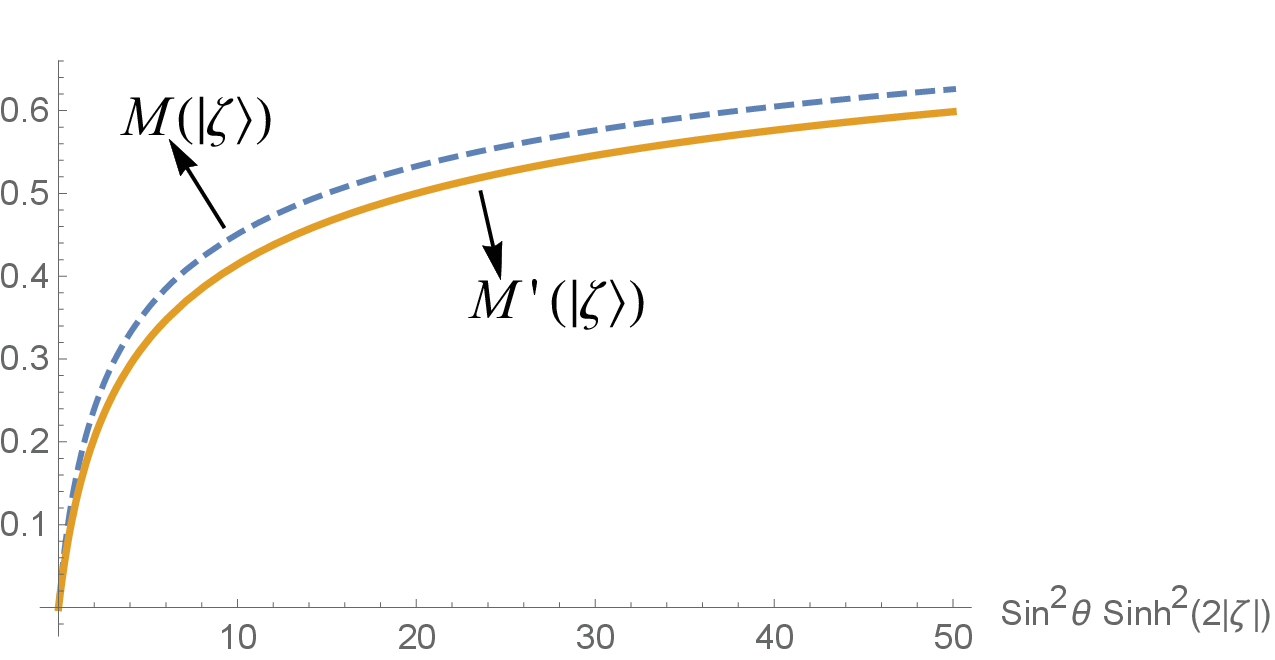}
\caption{$M(|\zeta \rangle )$ and $M'(|\zeta \rangle )$ versus $\sin ^{2}\theta
\sinh ^{2}$ in Eqs. (\ref{eq4-17},\ref{eq4-18}).}
\end{figure}

\section{Summary and outlook}

We established a resource theory of imaginarity for Gaussian states. To this
aim, under the Fock basis, we determined the real Gaussian states and real
Gaussian channels via the means and covariances of Gaussian states. We
provided two imaginarity measures based on the fidelity which all have closed expressions. As a
byproduct, we proved that the conjugate of a Gaussian state is still a
Gaussian state. We also discussed the imaginarity of some one-mode Gaussian
states.

There remained many open questions for future explorations. First, for
 the two imaginarity measures $M(\rho)$ and $M'(\rho)$ provided in this work, are there some physically operational interpretations linked to them? Second, does $M(\rho)\geq M'(\rho)$ hold for all Gaussian states? Third, are
there some other imaginarity measures for Gaussian states satisfying the
conditions (M1) and (M2) in this work? Lastly, the properties of state
conversions under real Gaussian channels are worthy of further
investigations.

\section*{ACKNOWLEDGMENTS}

This work was supported by the Natural Science Basic Research Plan in
Shaanxi Province of China (Program No. 2022JM-012). The author thanks Alessio Serafini, Shuanping Du and Kailiang Lin for helpful discussions.

\section*{Appendix A: Proof of Theorem 1}
\setcounter{equation}{0} \renewcommand\theequation{A\arabic{equation}}

We set three steps to prove Theorem 1.

(A.1). We first prove that if Gaussian state $\rho (\overline{X},V)$ is real
then $\{X_{2l}=0\}_{l=1}^{N}$ and $\{V_{2l-1,2m}=0\}_{l,m=1}^{N}$, this step is
comparatively straightforward.

Expand the Gaussian state $\rho (\overline{X},V)$ in the Fock basis $%
\{|j\rangle \}_{j}^{\otimes m}$ as
\begin{eqnarray}
\rho (\overline{X},V)=\sum_{j_{1},k_{1},...,j_{N},k_{N}=0}^{\infty }\rho
_{j_{1}k_{1},j_{2}k_{2},...,j_{N}k_{N}} \ \ \ \   \nonumber \\
\cdot|j_{1}\rangle \langle k_{1}|\otimes
|j_{2}\rangle \langle k_{2}|\otimes ...\otimes |j_{N}\rangle \langle k_{N}|, \label{eqA1}
\end{eqnarray}%
where $\rho _{j_{1}k_{1},j_{2}k_{2},...,j_{N}k_{N}}=\langle j_{1}|\langle
j_{2}|...\langle j_{N}|\rho |k_{1}\rangle |k_{2}\rangle ...|k_{N}\rangle \in
R$ for any $\{j_{1},k_{1},j_{2},k_{2},...j_{N},k_{N}\}\subset \{0,1,2,...\}.$
We also use the symbols $\rho _{j_{1}k_{1}}=\langle j_{1}|\rho
^{(1)}|k_{1}\rangle $ with $\rho ^{(1)}$ the reduced state of $\rho $ to the
first mode, $\rho _{j_{1}k_{1};j_{2}k_{2}}=\langle j_{1}|\langle j_{2}|\rho
^{(12)}|k_{1}\rangle |k_{2}\rangle $ with $\rho ^{(12)}$ the two-mode
reduced state of $\rho $ to the first and second modes.

Without loss of generality, we only need to prove that if $\rho (\overline{X}%
,V)$ is real then $\overline{X}_{2}=0,$ $V_{12}=V_{14}=V_{23}=0.$ Note that $%
\overline{X}_{1}$ is not on an equal footing with $\overline{X}_{2}$ by the
definition of Eqs. (\ref{eq2-4},\ref{eq2-5},\ref{eq2-12}), any $\overline{X}_{2l-1}$ ($\overline{X}_{2l}$) is on
an equal footing with $\overline{X}_{1}$ ($\overline{X}_{2}$). Similarly, $%
V_{12},$ $V_{14}$ and $V_{23}$ have distinct meanings by the definition of
Eqs. (\ref{eq2-4},\ref{eq2-5},\ref{eq2-13}), any $V_{2l-1,2m}$ has the similar situation with one of $%
\{V_{12},V_{14},V_{23}\}.$

From Eqs. (\ref{eqA1},\ref{eq2-1},\ref{eq2-2},\ref{eq2-4},\ref{eq2-5},\ref{eq2-12}), direct calculations show that
\begin{eqnarray}
\text{tr}(\rho \widehat{a}_{1}\rangle  &=&\sum_{j_{1}=0}^{\infty }\rho
_{j_{1},j_{1}+1}^{\ast }\sqrt{j_{1}+1},  \label{eqA2} \\
\text{tr}(\rho \widehat{a}_{1}^{\dag }) &=&[\text{tr}(\widehat{a}%
_{1})]^{\ast }=\sum_{j_{1}=0}^{\infty }\rho _{j_{1},j_{1}+1}\sqrt{j_{1}+1},  \label{eqA3}
\\
\overline{X}_{1} &=&\langle \widehat{a}_{1}+\widehat{a}_{1}^{\dag }\rangle ,  \label{eqA4}
\\
\overline{X}_{2} &=&-i\langle \widehat{a}_{1}-\widehat{a}_{1}^{\dag }\rangle.   \label{eqA5}
\end{eqnarray}%
We see that if $\rho (\overline{X},V)$ is real, then $\overline{X}_{2}=0.$

To express $V_{12},V_{14}$ and $V_{23},$ from Eqs. (\ref{eqA1},\ref{eq2-1},\ref{eq2-2},\ref{eq2-4},\ref{eq2-5},\ref{eq2-13}) we derive that
\begin{eqnarray}
\text{tr}(\rho \widehat{a}_{1}^{2})=\sum_{j_{1}=0}^{\infty }\rho
_{j_{1},j_{1}+2}^{\ast }\sqrt{(j_{1}+1)(j_{1}+2)}, \ \ \ \ \      \label{eqA6} \\
\text{tr}(\rho \widehat{a}_{1}^{\dag 2})=\sum_{j=0}^{\infty }\rho
_{j_{1},j_{1}+2}\sqrt{(j+1)(j+2)}, \ \ \ \ \ \ \ \     \label{eqA7}   \\
V_{12}=-i[\text{tr}(\rho \widehat{a}_{1}^{2})-\text{tr}(\rho \widehat{a}%
_{1}^{\dag 2})]-\overline{X}_{1}\overline{X}_{2}; \ \ \    \label{eqA8}  \\
\text{tr}(\rho \widehat{a}_{1}\widehat{a}_{2})=\sum_{j_{1}=0,j_{2}=0}^{\infty }\rho _{j_{1},j_{1}+1;j_{2},j_{2}+1}^{\ast
}\sqrt{(j_{1}+1)(j_{2}+1)},   \notag \\   \label{eqA9}  \\
\text{tr}(\rho \widehat{a}_{1}^{\dagger }\widehat{a}_{2}^{\dag })=\sum_{j_{1}=0,j_{2}=0}^{\infty }\rho _{j_{1},j_{1}+1;j_{2},j_{2}+1}\sqrt{%
(j_{1}+1)(j_{2}+1)},   \notag \\   \label{eqA10} \\
\text{tr}(\rho \widehat{a}_{1}\widehat{a}_{2}^{\dag })=\sum_{j_{1}=0,j_{2}=1}^{\infty }\rho _{j_{1},j_{1}+1;j_{2},j_{2}-1}^{\ast
}\sqrt{(j_{1}+1)j_{2}},  \ \ \ \ \ \ \     \label{eqA11}     \\
\text{tr}(\rho \widehat{a}_{1}^{\dagger }\widehat{a}_{2})=\sum_{j_{1}=0,j_{2}=1}^{\infty }\rho _{j_{1},j_{1}+1;j_{2},j_{2}-1}\sqrt{%
(j_{1}+1)j_{2}}, \ \ \ \ \ \ \     \label{eqA12}     \\
V_{14}=-i\text{tr}[\rho (\widehat{a}_{1}\widehat{a}_{2}-\widehat{a}%
_{1}^{\dag }\widehat{a}_{2}^{\dag }-\widehat{a}_{1}\widehat{a}_{2}^{\dag }+%
\widehat{a}_{1}^{\dag }\widehat{a}_{2})]-\overline{X}_{1}\overline{X}_{4}, \ \ \ \   \label{eqA13}  \\
V_{23}=-i\text{tr}[\rho (\widehat{a}_{1}\widehat{a}_{2}-\widehat{a}%
_{1}^{\dag }\widehat{a}_{2}^{\dag }+\widehat{a}_{1}\widehat{a}_{2}^{\dag }-%
\widehat{a}_{1}^{\dag }\widehat{a}_{2})]-\overline{X}_{2}\overline{X}_{3}. \ \ \ \  \label{eqA14}
\end{eqnarray}%
It follows that if $\rho (\overline{X},V)$ is real, then $%
V_{12}=V_{14}=V_{23}=0.$

(A.2). Next, we prove that for Gaussian state $\rho (\overline{X},V),$ if $%
\{X_{2l}=0\}_{l=1}^{N}$ and $\{V_{2l-1,2m}=0\}_{l,m=1}^{N},$ then $\rho (%
\overline{X},V)$ must be real. This step is comparatively difficult. Our
proof is inspired by Ref. \cite{PRA-Xu-2016} (see the Appendix A wherein).

For a quantum state $\rho $ in $\overline{H}^{\otimes N},$ its
characteristic function $\chi (\rho ,\xi )$ determines $\rho $ via the
relation \cite{book-2017-Serafini}
\begin{equation}
\rho =\int \frac{d^{2N}\xi }{\pi ^{N}}\chi (\rho ,\xi )D(-\xi ).   \label{eqA15}
\end{equation}
Then
\begin{eqnarray}
&&\langle j_{1}|\langle j_{2}|...\langle j_{N}|\rho |k_{1}\rangle
|k_{2}\rangle ...|k_{N}\rangle  \notag \\
&=&\int \frac{d^{2N}\lambda }{\pi ^{N}}\chi (\rho ,\xi )\langle
j_{1}|D(-\lambda _{1})|k_{1}\rangle ...\langle j_{N}|D(-\lambda
_{N})|k_{N}\rangle , \notag \\   \label{eqA16}
\end{eqnarray}%
where $\lambda _{1}=\xi _{1}+i\xi _{2},\lambda _{2}=\xi _{3}+i\xi _{4},$...,
$\lambda _{N}=\xi _{2N-1}+i\xi _{2N},$%
\begin{eqnarray}
D(\lambda _{1}) &=&D((\xi _{1},\xi _{2})^{T})=\exp (\lambda _{1}\widehat{a}%
_{1}^{\dagger }-\lambda _{1}^{\ast }\widehat{a}_{1}),   \label{eqA17} \\
D(\xi ) &=&D(\lambda _{1})D(\lambda _{2})...D(\lambda _{N}),   \label{eqA18}
\end{eqnarray}%
$D(\lambda _{2})=D((\xi _{3},\xi _{4})^{T}),$..., $D(\lambda _{N})=D((\xi
_{2N-1},\xi _{2N})^{T}).$ We further let $(\xi _{1},\xi _{2},\xi _{3},\xi
_{4},...,\xi _{2N-1},\xi _{2N})=(x_{\lambda _{1}},y_{\lambda
_{1}},x_{\lambda _{2}},y_{\lambda _{2}},...,x_{\lambda _{N}},y_{\lambda
_{N}}),$ $d^{2}\lambda _{1}=dx_{\lambda _{1}}dy_{\lambda _{1}},$ $%
d^{2N}\lambda =d^{2N}\xi \boldsymbol{=}dx_{\lambda _{1}}dy_{\lambda
_{1}}dx_{\lambda _{2}}dy_{\lambda _{2}}...dx_{\lambda _{1}}dy_{\lambda _{1}}.$
In Eq. (\ref{eqA16}),
\begin{eqnarray}
\langle j_{1}|D(-\lambda _{1})|k_{1}\rangle \ \ \ \ \ \ \ \ \  \ \ \ \ \ \   \ \ \ \ \ \  \ \ \ \ \ \   \ \ \ \ \ \ \ \ \ \ \ \    \notag \\
=\int \frac{d^{2}\alpha _{1}}{\pi }\frac{d^{2}\beta _{1}}{\pi }\langle
j_{1}|\alpha _{1}\rangle \langle \alpha _{1}|D(-\lambda _{1})|\beta
_{1}\rangle \langle \beta _{1}|k_{1}\rangle \ \ \  \label{eqA19} \\
=\int \frac{d^{2}\alpha _{1}}{\pi }\frac{d^{2}\beta _{1}}{\pi }\frac{%
\alpha _{1}^{j_{1}}\beta _{1}^{\ast k_{1}}}{\sqrt{j_{1}!k_{1}!}}\exp b_{1}, \ \ \ \ \ \ \ \ \ \  \ \ \ \ \ \  \ \ \ \ \ \   \label{eqA20} \\
b_{1}=-\frac{1}{2}(x_{\alpha _{1}},y_{\alpha _{1}},x_{\beta
_{1}},y_{\beta _{1}},x_{\lambda _{1}},y_{\lambda _{1}})Q_{1} \ \ \ \ \ \ \  \ \ \ \ \ \ \ \ \notag \\
(x_{\alpha
_{1}},y_{\alpha _{1}},x_{\beta _{1}},y_{\beta _{1}},x_{\lambda
_{1}},y_{\lambda _{1}})^{T}, \ \ \   \label{eqA21}  \\
Q_{1}=\left(
\begin{array}{cccccc}
2 & 0 & -1 & -i & 1 & i \\
0 & 2 & i & -1 & -i & 1 \\
-1 & i & 2 & 0 & -1 & i \\
-i & -1 & 0 & 2 & -i & -1 \\
1 & -i & -1 & -i & 1 & 0 \\
i & 1 & i & -1 & 0 & 1%
\end{array}
\right).   \ \ \ \ \ \  \ \ \ \ \ \ \ \ \ \ \  \label{eqA22}
\end{eqnarray}%
In Eq. (\ref{eqA19}), $\alpha _{1}=x_{\alpha _{1}}+iy_{\alpha _{1}},$ $\beta
_{1}=x_{\beta _{1}}+iy_{\beta _{1}},$ $\{x_{\alpha _{1}},y_{\alpha
_{1}},x_{\beta _{1}},y_{\beta _{1}}\}\subset R,$ $d^{2}\alpha
_{1}=dx_{\alpha _{1}}dy_{\alpha _{1}},$ $d^{2}\beta _{1}=dx_{\beta
_{1}}dy_{\beta _{1}}.$ Below we will use $\alpha _{2},\beta _{2},$..., $%
\alpha _{N},\beta _{N}$ and $d^{2N}\alpha =dx_{\alpha _{1}}dy_{\alpha
_{1}}...dx_{\alpha _{N}}dy_{\alpha _{N}},$ $d^{2N}\beta =dx_{\beta
_{1}}dy_{\beta _{1}}...dx_{\beta _{N}}dy_{\beta _{N}}$ similarly. In Eq. (\ref{eqA19}),
\begin{equation}
|\alpha _{1}\rangle =D(\alpha _{1})|0\rangle =e^{-\frac{|\alpha _{1}|^{2}}{2}%
}\sum_{j_{1}=0}^{\infty }\frac{\alpha _{1}^{j_{1}}}{\sqrt{j_{1}!}}%
|j_{1}\rangle  \label{eqA23}
\end{equation}%
is the Glauber coherent state, $|\beta _{1}\rangle $ similarly, and we have used the relations $\int \frac{d^{2}\alpha _{1}}{\pi }|\alpha _{1}\rangle \langle \alpha _{1}|=\int \frac{d^{2}\beta _{1}}{\pi }|\beta
_{1}\rangle \langle \beta _{1}|=I.$ In Eq. (\ref{eqA20}),
we have used $\langle \alpha _{1}|D(-\lambda _{1})|\beta _{1}\rangle
=\langle 0|D(-\alpha _{1})D(-\lambda _{1})D(\beta _{1})|0\rangle $ and the
relation
\begin{equation}
D(\alpha _{1})D(\beta _{1})=D(\alpha _{1}+\beta _{1})\exp \frac{\alpha
_{1}\beta _{1}^{\ast }-\alpha _{1}^{\ast }\beta _{1}}{2}.   \label{eqA24}
\end{equation}

Now we consider the case that $\rho $ is a Gaussian state $\rho =\rho (%
\overline{X},V).$ Taking Eq. (\ref{eqA20}) and the characteristic function $\chi (\rho ,\xi )$ in
Eq. (\ref{eq2-15}) into Eq. (\ref{eqA16}), we find
\begin{widetext}
\begin{eqnarray}
&&\langle j_{1}|\langle j_{2}|...\langle j_{N}|\rho |k_{1}\rangle
|k_{2}\rangle ...|k_{N}\rangle
=\int \frac{d^{2N}\lambda }{\pi ^{N}}\frac{d^{2N}\alpha }{\pi ^{N}}\frac{%
d^{2N}\beta }{\pi ^{N}}\frac{\alpha _{1}^{j_{1}}\beta _{1}^{\ast
k_{1}}...\alpha _{N}^{j_{N}}\beta _{N}^{\ast k_{N}}}{\sqrt{%
j_{1}!k_{1}!...j_{N}!k_{N}!}}\exp b_{2},   \label{eqA25}  \\
&&b_{2}=-\frac{1}{2}\Gamma ^{T}Q\Gamma +B^{\prime }\Gamma,  \ \ \ \ \ \ \ \ \ \ \  \ \ \ \ \ \ \ \ \ \ \ \ \ \ \ \ \ \ \ \ \ \ \ \ \ \ \ \ \ \ \\ \label{eqA26}
&&\Gamma=(x_{\alpha _{1}},y_{\alpha _{1}},x_{\beta _{1}},y_{\beta
_{1}},x_{\alpha _{2}},y_{\alpha _{2}},x_{\beta _{2}},y_{\beta _{2}},...,
x_{\alpha _{N}},y_{\alpha _{N}},x_{\beta _{N}},y_{\beta _{N}},x_{\lambda
_{1}},y_{\lambda _{1}},x_{\lambda _{2}},y_{\lambda _{2}},...x_{\lambda
_{N}},y_{\lambda _{N}})^{T},    \label{eqA27}  \\
&&B^{\prime }=(0,0,...,0,0,-i\overline{X}_{2},i\overline{X}_{1},-i\overline{X}%
_{4},i\overline{X}_{3},...,-i\overline{X}_{2N},i\overline{X}_{2N-1}),  \label{eqA28} \\
&&Q=\left(
\begin{array}{cccccccccccccc}
2 & 0 & -1 & -i & 0 & 0 & 0 & 0 & ... & 1 & i & 0 & 0 & ... \\
0 & 2 & i & -1 & 0 & 0 & 0 & 0 & ... & -i & 1 & 0 & 0 & ... \\
-1 & i & 2 & 0 & 0 & 0 & 0 & 0 & ... & -1 & i & 0 & 0 & ... \\
-i & -1 & 0 & 2 & 0 & 0 & 0 & 0 & ... & -i & -1 & 0 & 0 & ... \\
0 & 0 & 0 & 0 & 2 & 0 & -1 & -i & ... & 0 & 0 & 1 & i & ... \\
0 & 0 & 0 & 0 & 0 & 2 & i & -1 & ... & 0 & 0 & -i & 1 & ... \\
0 & 0 & 0 & 0 & -1 & i & 2 & 0 & ... & 0 & 0 & -1 & i & ... \\
0 & 0 & 0 & 0 & -i & -1 & 0 & 2 & ... & 0 & 0 & -i & -1 & ... \\
... & ... & ... & ... & ... & ... & ... & ... & ... & ... & ... & ... & ...
& ... \\
1 & -i & -1 & -i & 0 & 0 & 0 & 0 & ... & 1+V_{22} & -V_{21} & V_{24} &
-V_{23} & ... \\
i & 1 & i & -1 & 0 & 0 & 0 & 0 & ... & -V_{12} & 1+V_{11} & -V_{14} & V_{13}
& ... \\
0 & 0 & 0 & 0 & 1 & -i & -1 & -i & ... & V_{42} & -V_{41} & 1+V_{44} &
-V_{43} & ... \\
0 & 0 & 0 & 0 & i & 1 & i & -1 & ... & -V_{32} & V_{31} & -V_{34} & 1+V_{33}
& ... \\
... & ... & ... & ... & ... & ... & ... & ... & ... & ... & ... & ... & ...
& ...%
\end{array}%
\right).   \label{eqA29}
\end{eqnarray}
\end{widetext}

We introduce the Gaussian integral
\begin{equation}
J=\int d^{2N}\lambda d^{2N}\alpha d^{2N}\beta \exp
(b_{2}+\sum_{l=1}^{N}(u_{l}\alpha _{l}+v_{l}\beta _{l}^{\ast }),  \label{eqA30}
\end{equation}
where $\{u_{l},v_{l}\}_{l=1}^{N}\subset R,$ thus
\begin{eqnarray}
\langle j_{1}|\langle j_{2}|...\langle j_{N}|\rho |k_{1}\rangle
|k_{2}\rangle ...|k_{N}\rangle =\frac{1}{\pi ^{3N}\sqrt{j_{1}!k_{1}!...j_{N}!k_{N}!}} \notag \\
\cdot\left( \frac{\partial ^{j_{1}}}{\partial u_{1}^{j_{1}}}\frac{\partial ^{k_{1}}}{\partial
v_{1}^{k_{1}}}...\frac{\partial ^{j_{N}}}{\partial u_{N}^{j_{N}}}\frac{%
\partial ^{k_{N}}}{\partial v_{N}^{k_{N}}}J\right)
_{\{u_{l}=v_{l}=0\}_{l=1}^{N}}. \ \ \ \   \label{eqA31}
\end{eqnarray}
To calculate $J,$ we write
\begin{equation}
b_{2}+\sum_{l=1}^{N}(u_{l}\alpha _{l}+v_{l}\beta _{l}^{\ast })=-\frac{1}{2}%
\Gamma ^{T}Q\Gamma +B\Gamma ,  \label{eqA32}
\end{equation}
with
\begin{eqnarray}
B=(u_{1},iu_{1},v_{1},-iv_{1},u_{2},iu_{2},v_{2},-iv_{2},...,u_{N},iu_{N},v_{N},\notag \\
-iv_{N}, -i\overline{X}_{2},i\overline{X}_{1},-i\overline{X}_{4},i%
\overline{X}_{3},...,-i\overline{X}_{2N},i\overline{X}_{2N-1}). \ \ \ \  \label{eqA33}
\end{eqnarray}
Employing the Gaussian integral formula one gets
\begin{equation}
J=\frac{(2\pi )^{3N}}{\sqrt{\det Q}}\exp (\frac{1}{2}B^{T}Q^{-1}B).   \label{eqA34}
\end{equation}

Now we prove that if a Gaussian state $\rho (\overline{X},V)$ satisfies $%
\{X_{2l}=0\}_{l=1}^{N}$ and $\{V_{2l-1,2m}=0\}_{l,m=1}^{N},$ then $J\in R,$
hence Eq. (\ref{eqA31}) implies that $\rho (\overline{X},V)$ must be real. Observe that
if $\{\overline{X}_{2l}=0\}_{l=1}^{N}$ and $\{V_{2l-1,2m}=0\}_{l,m=1}^{N},$ then in Eq.
(\ref{eqA29}),
\begin{eqnarray}
\{Q_{2l-1,2m-1},Q_{2l,2m}\}_{l,m=1}^{N} &\subset &R,  \label{eqA35} \\
\{Q_{2l-1,2m},Q_{2l,2m-1}\}_{l,m=1}^{N} &\subset &iR,  \label{eqA36} \\
\{\overline{X}_{2l-1}\}_{l=1}^{N} \subset R, \ \
\{\overline{X}_{2l}\}_{l=1}^{N} &\subset &iR,  \label{eqA37} \\
iR=\{ix|x\in R\}.&&  \label{eqA38}
\end{eqnarray}
Let
\begin{equation}
Q=E^{\ast }Q^{\prime }E,  \label{eqA39}
\end{equation}
with
\begin{equation}
E=\text{diag}\{i,1,i,1,...,i,1\},  \label{eqA40}
\end{equation}
hence%
\begin{eqnarray}
E^{\ast }&=&E^{-1},  \label{eqA41} \\
Q_{2l-1,2m-1}^{\prime } &=&Q_{2l-1,2m-1},  \label{eqA42} \\
Q_{2l,2m}^{\prime }&=&Q_{2l,2m},  \label{eqA43} \\
Q_{2l-1,2m}^{\prime } &=&iQ_{2l-1,2m},  \label{eqA44} \\
Q_{2l,2m-1}^{\prime }&=&-iQ_{2l,2m-1}.  \label{eqA45} \\
B^{T}Q^{-1}B &=&B^{T}E^{\ast }Q^{\prime -1}EB. \label{eqA46}
\end{eqnarray}

We see that $Q^{\prime }$ is a real matrix, $\{(EB)_{l}\}_{l=1}^{N}\subset
iR,$ $\{(B^{T}E^{\ast })_{l}\}_{l=1}^{N}\subset iR.$ It follows that Eq. (\ref{eqA46})
is a real number and
\begin{equation}
\exp (\frac{1}{2}B^{T}A^{-1}B)\geq 0. \label{eqA47}
\end{equation}

Let $\{j_{1},k_{1},j_{2},k_{2},...j_{N},k_{N}\}$ all be zero, then
\begin{eqnarray}
\lbrack 0,1] &\ni &\langle 0|\langle 0|...\langle 0|\rho |0\rangle |0\rangle
...|0\rangle  \notag \\
&=&\frac{1}{\pi ^{3N}}J_{\{u_{l}=v_{l}=0\}_{l=1}^{N}}  \notag  \\
&=&\frac{2^{3N}}{\sqrt{\det Q}}\exp (\frac{1}{2}B^{T}Q^{-1}B)_{%
\{u_{l}=v_{l}=0\}_{l=1}^{N}}. \ \  \label{eqA48}
\end{eqnarray}
As a result,
\begin{equation}
\det Q>0.  \label{eqA49}
\end{equation}
Eqs. (\ref{eqA34},\ref{eqA31}) imply that $J\in R$ and $\rho (\overline{X},V)$ must be real.

(A.3). From the viewpoint of mathematical rigor, we need to prove that the
Gaussian integral in Eqs. (\ref{eqA30},\ref{eqA32},\ref{eqA33},\ref{eqA29},\ref{eqA27}) is convergent, then Eq. (\ref{eqA34}) is valid. To this aim, we prove $\text{Re}Q$
is positive definite (the convergence condition of Gaussian integral see for
example \cite{book-1989-Folland}).

Decompose $Q$ as
\begin{equation}
Q=Q_{2}+Q_{3}, \label{eqA50}
\end{equation}%
where $Q_{2}$ is obtained by deleting all $\{V_{lm}\}_{l,m=1}^{2N}$ in $Q.$
Since $V\succ 0,$ then
\begin{equation}
Q_{3}=\Omega V\Omega ^{T}\succ 0. \label{eqA51}
\end{equation}%
There exists a permutation matrix $P_{0}$ such that
\begin{equation}
Q_{2}=P_{0}(\oplus _{l=1}^{N}Q_{1})P_{0}^{T}, \label{eqA52}
\end{equation}%
where $P_{0}$ permutes the rows of $(\oplus _{l=1}^{N}Q_{1})$ while $%
P_{0}^{T}$ permutes the columns of $(\oplus _{l=1}^{N}Q_{1})$ in the same
way. Taking the real part of Eqs. (\ref{eqA50},\ref{eqA52}) gives
\begin{eqnarray}
\text{Re}Q &=&\text{Re}Q_{2}+Q_{3},  \label{eqA53} \\
\text{Re}Q_{2} &=&P_{0}(\oplus _{l=1}^{N}\text{Re}Q_{1})P_{0}^{T}. \label{eqA54}
\end{eqnarray}
$\text{Re}Q_{1}$ is symmetric and direct calculation shows that $\text{Re}Q_{1}$ has the eigenvalues $\{2+%
\sqrt{3},2+\sqrt{3},2-\sqrt{3},2-\sqrt{3},1,1\}$, thus Re$Q_{1}\succ 0,$ $%
\text{Re}Q_{2}\succ 0.$ Together with $Q_{3}=\Omega V\Omega ^{T}\succ 0,$
then we get $\text{Re}Q\succ 0.$

We then complete the proof of Theorem 1.

\section*{Appendix B: Proof of Theorem 2}
\setcounter{equation}{0} \renewcommand\theequation{B\arabic{equation}}

(B.1). For $N$-mode Gaussian state $\rho (\overline{X},V),$ we define the
real column vector $\overline{X}^{\prime }=(\overline{X}_{1}^{\prime },%
\overline{X}_{2}^{\prime },...,\overline{X}_{2N}^{\prime })^{T}$ and the $%
2N\times 2N$ real, symmetric matrix $V^{\prime }=(V_{lm}^{\prime
})_{l,m=1}^{2N}$ as
\begin{eqnarray}
\overline{X}_{l}^{\prime } &=&(-1)^{l+1}\overline{X}_{l},l\in \{1,2,...,2N\}, \label{eqB1}
\\
V_{lm}^{\prime } &=&(-1)^{l+m}V_{lm},l,m\in \{1,2,...,2N\}. \label{eqB2}
\end{eqnarray}

We first show that $(\overline{X}^{\prime },V^{\prime })$ determines a
Gaussian state $\rho ^{\prime }$ with $\overline{X}^{\prime }$ the mean and $%
V^{\prime }$ the covariance matrix. To do so, we need to prove that $%
V^{\prime }$ satisfies the uncertainty relation
\begin{equation}
V^{\prime }+i\Omega \succeq 0. \label{eqB3}
\end{equation}%
Since $V$ is the covariance matrix of Gaussian state $\rho (\overline{X},V),$
then the uncertainty relation
\begin{equation}
V+i\Omega \succeq 0  \label{eqB4}
\end{equation}%
holds. Taking the conjugate of the left-hand side of Eq. (\ref{eqB4}), gives
\begin{equation}
V-i\Omega \succeq 0. \label{eqB5}
\end{equation}%
That is, $V+i\Omega \succeq 0$ and $V-i\Omega \succeq 0$ are essentialy
equivalent.

Now introduce the matrix
\begin{equation}
O=\oplus _{l=1}^{N}\left(
\begin{array}{cc}
1 & 0 \\
0 & -1%
\end{array}%
\right).  \label{eqB6}
\end{equation}%
$O$ is a real orthogonal matrix and $O^{\dagger }=O.$ We can check that $%
OVO^{\dagger }=V^{\prime }$ and $O\Omega O^{\dagger }=-\Omega .$ Hence $%
O(V-i\Omega )O^{\dagger }\succeq 0$ and Eq. (\ref{eqB3}) follows.

(B.2). We prove that the conjugate of $\rho (\overline{X},V),$ $\rho ^{\ast
},$ has the mean $\overline{X}^{\prime }$ and covariance matrix $V^{\prime}. $

Similar to Eqs. (\ref{eqA6}-\ref{eqA14}), we have
\begin{eqnarray}
\text{tr}(\rho \widehat{a}_{1}^{\dagger }\widehat{a}_{1})
=\sum_{j=0}^{\infty }j_{1}\rho _{j_{1}j_{1}},  \ \ \ \ \  \ \ \ \ \  \ \ \ \ \  \ \ \ \ \  \ \ \ \ \  \ \ \ \ \  \label{eqB7} \\
V_{11}=[\text{tr}(\rho (\widehat{a}_{1}^{2}+\widehat{a}_{1}^{\dag 2}+2%
\widehat{a}_{1}^{\dagger }\widehat{a}_{1}+1)]-\overline{X}_{1}^{2},  \ \ \ \ \  \ \ \ \ \  \ \  \ \ \ \ \label{eqB8} \\
V_{22}=-[\text{tr}(\rho (\widehat{a}_{1}^{2}+\widehat{a}_{1}^{\dag 2}-2%
\widehat{a}_{1}^{\dagger }\widehat{a}_{1}-1)]-\overline{X}_{2}^{2},  \ \ \ \ \  \ \ \ \ \   \ \ \ \label{eqB9}  \\
V_{13}=\text{tr}[\rho (\widehat{a}_{1}\widehat{a}_{2}+\widehat{a}%
_{1}^{\dag }\widehat{a}_{2}^{\dag }+\widehat{a}_{1}\widehat{a}_{2}^{\dag }+%
\widehat{a}_{1}^{\dag }\widehat{a}_{2})]-\overline{X}_{1}\overline{X}_{3}, \ \ \ \ \  \label{eqB10}  \\
V_{24}=-\text{tr}[\rho (\widehat{a}_{1}\widehat{a}_{2}+\widehat{a}%
_{1}^{\dag }\widehat{a}_{2}^{\dag }-\widehat{a}_{1}\widehat{a}_{2}^{\dag }-%
\widehat{a}_{1}^{\dag }\widehat{a}_{2})]-\overline{X}_{2}\overline{X}_{4}. \ \ \ \label{eqB11}
\end{eqnarray}

We see that if replace $\rho $ by its conjugate $\rho ^{\ast },$ then $(%
\overline{X},V)$ become $(\overline{X}^{\prime },V^{\prime }).$ This says
that the mean and covariance matrix of $\rho ^{\ast }$ are $(\overline{X}%
^{\prime },V^{\prime }).$

(B.3). Lastly, we prove that the Gaussian state $\rho ^{\prime }(\overline{X}%
^{\prime },V^{\prime })$ is just $\rho ^{\ast }.$ To this aim, we calculate
the matrix elements $\langle j_{1}|\langle j_{2}|...\langle j_{N}|\rho
^{\prime }|k_{1}\rangle |k_{2}\rangle ...|k_{N}\rangle $ of $\rho ^{\prime }(%
\overline{X}^{\prime },V^{\prime })$ in the Fock basis $\{|j\rangle
\}_{j}^{\otimes N}$. We need to show that
\begin{eqnarray}
&&\langle j_{1}|\langle j_{2}|...\langle j_{N}|\rho ^{\prime }|k_{1}\rangle
|k_{2}\rangle ...|k_{N}\rangle  \notag \\
&=&\langle j_{1}|\langle j_{2}|...\langle j_{N}|\rho ^{\ast }|k_{1}\rangle
|k_{2}\rangle ...|k_{N}\rangle  \\
&=&\langle k_{1}|\langle k_{2}|...\langle k_{N}|\rho |j_{1}\rangle
|j_{2}\rangle ...|j_{N}\rangle
\end{eqnarray}%
for any $\{j_{1},k_{1},j_{2},k_{2},...j_{N},k_{N}\}\subset \{0,1,2,...\}.$

One can check that when replacing $(\overline{X},V)$ by $(\overline{X}%
^{\prime },V^{\prime })$ and replacing $%
(u_{1},v_{1};u_{2},v_{2};...;u_{N},v_{N})$ by $%
(v_{1},u_{1};v_{2},u_{2};...;v_{N},u_{N})$ in Eqs. (\ref{eqA30},\ref{eqA32},\ref{eqA33},\ref{eqA27},\ref{eqA29}), the integral $J$
remains invariant. Together with Eq. (\ref{eqA31}), we then obtain $\rho ^{\prime }(%
\overline{X}^{\prime },V^{\prime })=\rho ^{\ast }.$

\section*{Appendix C: Proof of Theorem 3}
\setcounter{equation}{0} \renewcommand\theequation{C\arabic{equation}}

(C.1). One-mode case.

Consider the real Gaussian state $\rho (\overline{X},V)$ with
\begin{equation}
V=\left(
\begin{array}{cc}
V_{11} & 0 \\
0 & V_{22}%
\end{array}%
\right) ,\overline{X}=\left(
\begin{array}{c}
\overline{X}_{1} \\
0%
\end{array}%
\right).  \label{eqC1}
\end{equation}
Suppose $\phi =(d,T,N)$ is a real Gaussian channel,
\begin{equation}
T=\left(
\begin{array}{cc}
T_{11} & T_{12} \\
T_{21} & T_{22}%
\end{array}%
\right) ,N=\left(
\begin{array}{cc}
N_{11} & N_{12} \\
N_{12} & N_{22}%
\end{array}%
\right) ,d=\left(
\begin{array}{c}
d_{1} \\
d_{2}%
\end{array}%
\right) .  \ \ \ \  \label{eqC2}
\end{equation}
Eq. (\ref{eq3-1}) yields
\begin{equation}
T\overline{X}+d=\left(
\begin{array}{c}
T_{11}\overline{X}_{1}+d_{1} \\
T_{21}\overline{X}_{1}+d_{2}%
\end{array}%
\right).    \label{eqC3}
\end{equation}%
Varying $\overline{X}_{1}\in R,$ hence $T_{21}\overline{X}_{1}+d_{2}=0$
implies $T_{21}=d_{2}=0.$ Further,
\begin{eqnarray}
&&TVT^{T}+N   \nonumber \\
&=&\left(
\begin{array}{cc}
T_{11}^{2}V_{11}+T_{12}^{2}V_{22}+N_{11} & T_{12}T_{22}V_{22}+N_{12} \\
T_{12}T_{22}V_{22}+N_{12} & T_{22}^{2}V_{22}+N_{22}%
\end{array}%
\right).  \ \ \ \ \ \    \label{eqC4}
\end{eqnarray}%
Varying $V_{22}\in R,$ hence $%
T_{12}T_{22}V_{22}+N_{12}=T_{12}T_{22}V_{22}+N_{12}=0$ yield $%
T_{12}T_{22}=N_{12}=0.$ Then Theorem 2 holds for one-mode case.

(C.2). $N$-mode case.

Define the $2N\times 2N$ permutation matrix $P$ as
\begin{equation}
P_{l,2l-1}=P_{N+l,2l}=1\text{ for }l\in \{1,2,...,N\},  \label{eqC5}
\end{equation}%
and other elements are all zero. $P$ reorders the indices $(1,2,3,...,2N)^{T}
$ to $(1,3,5,...,2N-1,2,4,...,2N)^{T}.$ Consider the real Gaussian state $%
\rho (\overline{X},V)$, we find
\begin{eqnarray}
P\overline{X} &=&(\overline{X}_{1},\overline{X}_{3},...,\overline{X}%
_{2N-1},0,...,0)^{T},  \label{eqC6} \\
PVP^{T} &=&\left(
\begin{array}{cccccc}
V_{11} & V_{13} & ... & 0 & 0 & ... \\
V_{31} & V_{33} & ... & 0 & 0 & ... \\
... & ... & ... & ... & ... & ... \\
0 & 0 & ... & V_{22} & V_{24} & ... \\
0 & 0 & ... & V_{42} & V_{44} & ... \\
... & ... & ... & ... & ... & ...%
\end{array}%
\right)  \nonumber  \\
&=&\left(
\begin{array}{cc}
V_{1} & 0 \\
0 & V_{4}%
\end{array}%
\right) .   \label{eqC7}
\end{eqnarray}%
with $V_{1}=V_{o}$ and $V_{4}=V_{e}$ all $N\times N$ real matrices.

Suppose $\phi =(d,T,N)$ is a real Gaussian channel. $PTP^{T}$ and $PNP^{T}$
have the reordered structures similar to $PVP^{T}.$ We write
\begin{eqnarray}
PTP^{T}=\left(
\begin{array}{cc}
T_{1} & T_{2} \\
T_{3} & T_{4}%
\end{array}%
\right) ,PNP^{T}=\left(
\begin{array}{cc}
N_{1} & N_{2} \\
N_{2}^{T} & N_{4}%
\end{array}%
\right) ,  \label{eqC8} \\
Pd=(d_{1},d_{3},...,d_{2N-1},d_{2},d_{4},...,d_{2N})^{T},  \label{eqC9}
\end{eqnarray}
with $T_{1},T_{2},T_{3},T_{4},N_{1},N_{2}$ and $N_{3}$ all $N\times N$ real
matrices. Eq. (\ref{eq3-1}) yields
\begin{equation}
T\overline{X}+d=P^{T}[(PTP^{T})(P\overline{X})+(Pd)],  \label{eqC10}
\end{equation}%
Varying $\{\overline{X}_{2l-1}\}_{l=1}^{N}\subset R,$ we get $T_{3}=0$ and $%
d_{2l}=0$ for $l\in \{1,2,...,N\}.$ Eq. (\ref{eq3-1}) further yields
\begin{eqnarray}
&&TVT^{T}+N  \nonumber  \\
&=&P^{T}[(PTP^{T})(PVP^{T})(PT^{T}P^{T})+(PNP^{T})]P  \nonumber  \\
&=&P^{T}\left(
\begin{array}{cc}
T_{1}V_{1}T_{1}^{T}+T_{2}V_{4}T_{2}^{T}+N_{1} & T_{2}V_{4}T_{4}^{T}+N_{2}  \\
T_{4}V_{4}T_{2}^{T}+N_{2}^{T} & T_{4}V_{4}T_{4}^{T}+N_{4}%
\end{array}%
\right) P,  \nonumber  \\
&&T_{2}V_{4}T_{4}^{T}+N_{2}=0.  \label{eqC11}
\end{eqnarray}
Varying $\{V_{2l,2m}\}_{l,m=1}^{N}\subset R$ in Eq. (\ref{eqC11}), we get $N_{2}=0$, $%
T_{2}=0$ or $T_{4}=0.$ Then Theorem 3 follows for $N$-mode case.

\section*{Appendix D: Proof of Theorem 4}
\setcounter{equation}{0} \renewcommand\theequation{D\arabic{equation}}

Suppose $\phi =(d,T,N)$ is an $N$-mode real Gaussian channel and $\rho (%
\overline{X},V)$ is any $N$-mode Gaussian state. $P,$ $P\overline{X},$ $PTP^{T}$, $PNP^{T}$ and $Pd$
are defined similarly to Eqs. (\ref{eqC5},\ref{eqC6},\ref{eqC8},\ref{eqC9}). We also denote
\begin{eqnarray}
\overline{X}_{o} &=&(\overline{X}_{1},\overline{X}_{3},\overline{X}_{5},...,%
\overline{X}_{2N-1})^{T}, \label{eqD1} \\
\overline{X}_{e} &=&(\overline{X}_{2},\overline{X}_{4},\overline{X}_{6},...,%
\overline{X}_{2N})^{T}, \label{eqD2} \\
PVP^{T} &=&\left(
\begin{array}{cc}
V_{1} & V_{2} \label{eqD3} \\
V_{2}^{T} & V_{4}%
\end{array}%
\right) , \\
d_{o} &=&(d_{1},d_{3},d_{5},...,d_{2N-1})^{T}, \label{eqD4}
\end{eqnarray}
where $\{V_{2},V_{1}=V_{o},V_{4}=V_{e}\}$ are all $N\times N$ matrices.

(D.1). If $\phi =(d,T,N)$ is a completely real Gaussian channel, then
\begin{equation}
PTP^{T}=\left(
\begin{array}{cc}
T_{1} & T_{2} \\
0 & 0%
\end{array}%
\right) ,PNP^{T}=\left(
\begin{array}{cc}
N_{1} & 0 \\
0 & N_{4}%
\end{array}%
\right) .  \label{eqD5}
\end{equation}
We calculate the mean and covariance of $\phi \lbrack \rho (\overline{X},V)].
$ Eq. (\ref{eq3-1}) yield
\begin{eqnarray}
&&T\overline{X}+d=P^{T}[(PTP^{T})(P\overline{X})+(Pd)] \notag \\
&& \ \ \ \ \ \ \ \ \ \ \ =P^{T}\left(
\begin{array}{c}
T_{1}\overline{X}_{o}+T_{2}\overline{X}_{e}+d_{o} \\
0%
\end{array}%
\right),  \label{eqD6} \\
&&TVT^{T}+N  \notag \\
&=&P^{T}[(PTP^{T})(PVP^{T})(PT^{T}P^{T})+(PNP^{T})]P  \notag \\
&=&P^{T}\left(
\begin{array}{cc}
V'_{1}& 0 \\
0 & N_{4}%
\end{array}%
\right) P,  \label{eqD7} \\
V'_{1}&=&T_{1}V_{1}T_{1}^{T}+T_{1}V_{2}T_{2}^{T}+T_{2}V_{2}^{T}T_{1}^{T}+T_{2}V_{4}T_{2}^{T}+N_{1}.  \notag
\end{eqnarray}
Consequently, $\phi \lbrack \rho (\overline{X},V)]$ is a real Gaussian state.

(D.2). If $\phi =(d,T,N)$ is a covariant real Gaussian channel, then
\begin{equation}
PTP^{T}=\left(
\begin{array}{cc}
T_{1} & 0 \\
0 & T_{4}%
\end{array}%
\right) ,PNP^{T}=\left(
\begin{array}{cc}
N_{1} & 0 \\
0 & N_{4}%
\end{array}%
\right) .  \label{eqD8}
\end{equation}

We calculate the mean and covariance of $\phi \lbrack \rho (\overline{X},V)].$ Eq. (\ref{eq3-1}) yield
\begin{eqnarray}
&&T\overline{X}+d=P^{T}[(PTP^{T})(P\overline{X})+(Pd)] \notag  \\
&& \ \ \ \ \ \ \ \ \ \ \ =P^{T}\left(
\begin{array}{c}
T_{1}\overline{X}_{o}+d_{o} \\
T_{4}\overline{X}_{e}%
\end{array}%
\right),   \label{eqD9} \\
&&TVT^{T}+N \notag \\
&=&P^{T}[(PTP^{T})(PVP^{T})(PT^{T}P^{T})+(PNP^{T})]P \notag \\
&=&P^{T}\left(
\begin{array}{cc}
T_{1}V_{1}T_{1}^{T}+N_{1} & T_{1}V_{2}T_{4}^{T} \\
T_{4}V_{2}^{T}T_{1}^{T} & T_{4}V_{4}T_{4}^{T}+N_{4}%
\end{array}%
\right) P.   \label{eqD10}
\end{eqnarray}

Applying Theorem 2, $\{\phi \lbrack \rho (V,\overline{X})]\}^{\ast }$ is
still a Gaussian state with the mean and covariance matrix
\begin{eqnarray}
&&P^{T}\left(
\begin{array}{c}
T_{1}\overline{X}_{o}+d_{o} \\
-T_{4}\overline{X}_{e}%
\end{array}%
\right),  \label{eqD11} \\
&&P^{T}\left(
\begin{array}{cc}
T_{1}V_{1}T_{1}^{T}+N_{1} & -T_{1}V_{2}T_{4}^{T} \\
-T_{4}V_{2}^{T}T_{1}^{T} & T_{4}V_{4}T_{4}^{T}+N_{4}%
\end{array}%
\right)P.  \label{eqD12}
\end{eqnarray}
We see that Eqs. (\ref{eqD11},\ref{eqD12}) are just the mean and covariance matrix of $\phi \lbrack
\rho ^{\ast }(V^{\prime },\overline{X}^{\prime })],$ that is, $\overline{X}_{e}\rightarrow -\overline{X}_{e}$ and $V_{2}\rightarrow -V_{2}.$  Thus Eq. (\ref{eq3-7}) holds.

The proof of Eq. (\ref{eq3-8}) is similar to the proof of Eq. (\ref{eq3-7}).

\section*{Appendix E: Proof of Theorem 5}
\setcounter{equation}{0} \renewcommand\theequation{E\arabic{equation}}

We prove that Eq. (\ref{eq4-1}) fulfills (M1) and (M2). Eq. (\ref{eq4-1}) fulfilling (M1) is
apparent since for any two quantum states $\rho $ and $\sigma ,$ the
fidelity $F(\rho ,\sigma )$ $\geq 0$ and $F(\rho ,\sigma )$ $=0$ if and only
if $\rho =\sigma $ \cite{Nielsen-2010-quantum}. Now we prove Eq. (\ref{eq4-1}) fulfills (M2).

For a real Gaussian channel $\phi ,$ if $\phi $ is a completely real
Gaussian channel, then $\phi (\rho )$ is a real Gaussian state and $M(\phi
(\rho ))=0\leq M(\rho )$ for any Gaussian state $\rho .$

For a real Gaussian channel $\phi ,$ if $\phi $ is a covariant real Gaussian
channel, then from Theorem 4 we have $[\phi (\rho )]^{\ast }=\phi (\rho
^{\ast }),$ and
\begin{eqnarray}
M(\phi (\rho )) &=&1-F(\phi (\rho ),(\phi (\rho ))^{\ast }) \notag \\
&=&1-F(\phi (\rho ),\phi (\rho ^{\ast }))  \notag \\
&\leq &1-F(\rho ,\rho ^{\ast })=M(\rho ).  \label{E1}
\end{eqnarray}%
In the inequality we have used the monotonicity of the fidelity under a
quantum channel $\phi $, $F(\phi (\rho ),\phi (\sigma ))\geq F(\rho ,\sigma )
$ for any two quantum states $\rho $ and $\sigma $ \cite{Nielsen-2010-quantum}.

The proof of Eq. (\ref{eq4-2}) fulfilling  (M1) and (M2) is similar to Eq. (\ref{eq4-1}). Theorem 5 then follows.


\begin{thebibliography}{50}%
\makeatletter
\providecommand \@ifxundefined [1]{%
 \@ifx{#1\undefined}
}%
\providecommand \@ifnum [1]{%
 \ifnum #1\expandafter \@firstoftwo
 \else \expandafter \@secondoftwo
 \fi
}%
\providecommand \@ifx [1]{%
 \ifx #1\expandafter \@firstoftwo
 \else \expandafter \@secondoftwo
 \fi
}%
\providecommand \natexlab [1]{#1}%
\providecommand \enquote  [1]{``#1''}%
\providecommand \bibnamefont  [1]{#1}%
\providecommand \bibfnamefont [1]{#1}%
\providecommand \citenamefont [1]{#1}%
\providecommand \href@noop [0]{\@secondoftwo}%
\providecommand \href [0]{\begingroup \@sanitize@url \@href}%
\providecommand \@href[1]{\@@startlink{#1}\@@href}%
\providecommand \@@href[1]{\endgroup#1\@@endlink}%
\providecommand \@sanitize@url [0]{\catcode `\\12\catcode `\$12\catcode
  `\&12\catcode `\#12\catcode `\^12\catcode `\_12\catcode `\%12\relax}%
\providecommand \@@startlink[1]{}%
\providecommand \@@endlink[0]{}%
\providecommand \url  [0]{\begingroup\@sanitize@url \@url }%
\providecommand \@url [1]{\endgroup\@href {#1}{\urlprefix }}%
\providecommand \urlprefix  [0]{URL }%
\providecommand \Eprint [0]{\href }%
\providecommand \doibase [0]{http://dx.doi.org/}%
\providecommand \selectlanguage [0]{\@gobble}%
\providecommand \bibinfo  [0]{\@secondoftwo}%
\providecommand \bibfield  [0]{\@secondoftwo}%
\providecommand \translation [1]{[#1]}%
\providecommand \BibitemOpen [0]{}%
\providecommand \bibitemStop [0]{}%
\providecommand \bibitemNoStop [0]{.\EOS\space}%
\providecommand \EOS [0]{\spacefactor3000\relax}%
\providecommand \BibitemShut  [1]{\csname bibitem#1\endcsname}%
\let\auto@bib@innerbib\@empty
\bibitem [{\citenamefont {Hickey}\ and\ \citenamefont
  {Gour}(2018)}]{JPA-Gour-2018}%
  \BibitemOpen
  \bibfield  {author} {\bibinfo {author} {\bibfnamefont {A.}~\bibnamefont
  {Hickey}}\ and\ \bibinfo {author} {\bibfnamefont {G.}~\bibnamefont {Gour}},\
  }\bibinfo {title} {Quantifying the imaginarity of quantum mechanics},\ \href
  {\doibase 10.1088/1751-8121/aabe9c} {\bibfield  {journal} {\bibinfo
  {journal} {Journal of Physics A: Mathematical and Theoretical}\ }\textbf
  {\bibinfo {volume} {51}},\ \bibinfo {pages} {414009} (\bibinfo {year}
  {2018})}\BibitemShut {NoStop}%
\bibitem [{\citenamefont {Wu}\ \emph {et~al.}(2021{\natexlab{a}})\citenamefont
  {Wu}, \citenamefont {Kondra}, \citenamefont {Rana}, \citenamefont {Scandolo},
  \citenamefont {Xiang}, \citenamefont {Li}, \citenamefont {Guo},\ and\
  \citenamefont {Streltsov}}]{PRL-Guo-2021}%
  \BibitemOpen
  \bibfield  {author} {\bibinfo {author} {\bibfnamefont {K.-D.}\ \bibnamefont
  {Wu}}, \bibinfo {author} {\bibfnamefont {T.~V.}\ \bibnamefont {Kondra}},
  \bibinfo {author} {\bibfnamefont {S.}~\bibnamefont {Rana}}, \bibinfo {author}
  {\bibfnamefont {C.~M.}\ \bibnamefont {Scandolo}}, \bibinfo {author}
  {\bibfnamefont {G.-Y.}\ \bibnamefont {Xiang}}, \bibinfo {author}
  {\bibfnamefont {C.-F.}\ \bibnamefont {Li}}, \bibinfo {author} {\bibfnamefont
  {G.-C.}\ \bibnamefont {Guo}}, \ and\ \bibinfo {author} {\bibfnamefont
  {A.}~\bibnamefont {Streltsov}},\ }\bibinfo {title} {Operational resource
  theory of imaginarity},\ \href {\doibase 10.1103/PhysRevLett.126.090401}
  {\bibfield  {journal} {\bibinfo  {journal} {Phys. Rev. Lett.}\ }\textbf
  {\bibinfo {volume} {126}},\ \bibinfo {pages} {090401} (\bibinfo {year}
  {2021}{\natexlab{a}})}\BibitemShut {NoStop}%
\bibitem [{\citenamefont {Wu}\ \emph {et~al.}(2021{\natexlab{b}})\citenamefont
  {Wu}, \citenamefont {Kondra}, \citenamefont {Rana}, \citenamefont {Scandolo},
  \citenamefont {Xiang}, \citenamefont {Li}, \citenamefont {Guo},\ and\
  \citenamefont {Streltsov}}]{PRA-Guo-2021}%
  \BibitemOpen
  \bibfield  {author} {\bibinfo {author} {\bibfnamefont {K.-D.}\ \bibnamefont
  {Wu}}, \bibinfo {author} {\bibfnamefont {T.~V.}\ \bibnamefont {Kondra}},
  \bibinfo {author} {\bibfnamefont {S.}~\bibnamefont {Rana}}, \bibinfo {author}
  {\bibfnamefont {C.~M.}\ \bibnamefont {Scandolo}}, \bibinfo {author}
  {\bibfnamefont {G.-Y.}\ \bibnamefont {Xiang}}, \bibinfo {author}
  {\bibfnamefont {C.-F.}\ \bibnamefont {Li}}, \bibinfo {author} {\bibfnamefont
  {G.-C.}\ \bibnamefont {Guo}}, \ and\ \bibinfo {author} {\bibfnamefont
  {A.}~\bibnamefont {Streltsov}},\ }\bibinfo {title} {Resource theory of
  imaginarity: Quantification and state conversion},\ \href {\doibase
  10.1103/PhysRevA.103.032401} {\bibfield  {journal} {\bibinfo  {journal}
  {Phys. Rev. A}\ }\textbf {\bibinfo {volume} {103}},\ \bibinfo {pages}
  {032401} (\bibinfo {year} {2021}{\natexlab{b}})}\BibitemShut {NoStop}%
\bibitem [{\citenamefont {Renou}\ \emph {et~al.}(2021)\citenamefont {Renou},
  \citenamefont {Trillo}, \citenamefont {Weilenmann}, \citenamefont {Le},
  \citenamefont {Tavakoli}, \citenamefont {Gisin}, \citenamefont {Ac{\'\i}n},\
  and\ \citenamefont {Navascu{\'e}s}}]{Nature-2021-Acin}%
  \BibitemOpen
  \bibfield  {author} {\bibinfo {author} {\bibfnamefont {M.-O.}\ \bibnamefont
  {Renou}}, \bibinfo {author} {\bibfnamefont {D.}~\bibnamefont {Trillo}},
  \bibinfo {author} {\bibfnamefont {M.}~\bibnamefont {Weilenmann}}, \bibinfo
  {author} {\bibfnamefont {T.~P.}\ \bibnamefont {Le}}, \bibinfo {author}
  {\bibfnamefont {A.}~\bibnamefont {Tavakoli}}, \bibinfo {author}
  {\bibfnamefont {N.}~\bibnamefont {Gisin}}, \bibinfo {author} {\bibfnamefont
  {A.}~\bibnamefont {Ac{\'\i}n}}, \ and\ \bibinfo {author} {\bibfnamefont
  {M.}~\bibnamefont {Navascu{\'e}s}},\ }\bibinfo {title} {Quantum theory based
  on real numbers can be experimentally falsified},\ \href
  {https://www.nature.com/articles/s41586-021-04160-4} {\bibfield  {journal}
  {\bibinfo  {journal} {Nature}\ }\textbf {\bibinfo {volume} {600}},\ \bibinfo
  {pages} {625} (\bibinfo {year} {2021})}\BibitemShut {NoStop}%
\bibitem [{\citenamefont {Zhu}(2021)}]{PRR-2021-Zhu}%
  \BibitemOpen
  \bibfield  {author} {\bibinfo {author} {\bibfnamefont {H.}~\bibnamefont
  {Zhu}},\ }\bibinfo {title} {Hiding and masking quantum information in complex
  and real quantum mechanics},\ \href {\doibase
  10.1103/PhysRevResearch.3.033176} {\bibfield  {journal} {\bibinfo  {journal}
  {Phys. Rev. Res.}\ }\textbf {\bibinfo {volume} {3}},\ \bibinfo {pages}
  {033176} (\bibinfo {year} {2021})}\BibitemShut {NoStop}%
\bibitem [{\citenamefont {Zhang}\ \emph {et~al.}(2021)\citenamefont {Zhang},
  \citenamefont {Hou}, \citenamefont {Li}, \citenamefont {Zhu}, \citenamefont
  {Xiang}, \citenamefont {Li},\ and\ \citenamefont {Guo}}]{PRAP-2021-Guo}%
  \BibitemOpen
  \bibfield  {author} {\bibinfo {author} {\bibfnamefont {R.-Q.}\ \bibnamefont
  {Zhang}}, \bibinfo {author} {\bibfnamefont {Z.}~\bibnamefont {Hou}}, \bibinfo
  {author} {\bibfnamefont {Z.}~\bibnamefont {Li}}, \bibinfo {author}
  {\bibfnamefont {H.}~\bibnamefont {Zhu}}, \bibinfo {author} {\bibfnamefont
  {G.-Y.}\ \bibnamefont {Xiang}}, \bibinfo {author} {\bibfnamefont {C.-F.}\
  \bibnamefont {Li}}, \ and\ \bibinfo {author} {\bibfnamefont {G.-C.}\
  \bibnamefont {Guo}},\ }\bibinfo {title} {Experimental masking of real quantum
  states},\ \href {\doibase 10.1103/PhysRevApplied.16.024052} {\bibfield
  {journal} {\bibinfo  {journal} {Phys. Rev. Appl.}\ }\textbf {\bibinfo
  {volume} {16}},\ \bibinfo {pages} {024052} (\bibinfo {year}
  {2021})}\BibitemShut {NoStop}%
\bibitem [{\citenamefont {Xue}\ \emph {et~al.}(2021)\citenamefont {Xue},
  \citenamefont {Guo}, \citenamefont {Li}, \citenamefont {Ye},\ and\
  \citenamefont {Li}}]{QIP-2021-Li}%
  \BibitemOpen
  \bibfield  {author} {\bibinfo {author} {\bibfnamefont {S.}~\bibnamefont
  {Xue}}, \bibinfo {author} {\bibfnamefont {J.}~\bibnamefont {Guo}}, \bibinfo
  {author} {\bibfnamefont {P.}~\bibnamefont {Li}}, \bibinfo {author}
  {\bibfnamefont {M.}~\bibnamefont {Ye}}, \ and\ \bibinfo {author}
  {\bibfnamefont {Y.}~\bibnamefont {Li}},\ }\bibinfo {title} {Quantification of
  resource theory of imaginarity},\ \href
  {https://link.springer.com/article/10.1007/s11128-021-03324-5} {\bibfield
  {journal} {\bibinfo  {journal} {Quantum Information Processing}\ }\textbf
  {\bibinfo {volume} {20}},\ \bibinfo {pages} {1} (\bibinfo {year}
  {2021})}\BibitemShut {NoStop}%
\bibitem [{\citenamefont {Kondra}\ \emph {et~al.}(2022)\citenamefont {Kondra},
  \citenamefont {Datta},\ and\ \citenamefont
  {Streltsov}}]{arxiv-2022-Streltsov}%
  \BibitemOpen
  \bibfield  {author} {\bibinfo {author} {\bibfnamefont {T.~V.}\ \bibnamefont
  {Kondra}}, \bibinfo {author} {\bibfnamefont {C.}~\bibnamefont {Datta}}, \
  and\ \bibinfo {author} {\bibfnamefont {A.}~\bibnamefont {Streltsov}},\
  }\bibinfo {title} {Real quantum operations and state transformations},\ \href
  {https://arxiv.org/abs/2210.15820v1} {\bibfield  {journal} {\bibinfo
  {journal} {arXiv preprint arXiv:2210.15820}\ } (\bibinfo {year}
  {2022})}\BibitemShut {NoStop}%
\bibitem [{\citenamefont {Li}\ \emph {et~al.}(2022{\natexlab{a}})\citenamefont
  {Li}, \citenamefont {Mao}, \citenamefont {Weilenmann}, \citenamefont
  {Tavakoli}, \citenamefont {Chen}, \citenamefont {Feng}, \citenamefont {Yang},
  \citenamefont {Renou}, \citenamefont {Trillo}, \citenamefont {Le},
  \citenamefont {Gisin}, \citenamefont {Ac\'{\i}n}, \citenamefont
  {Navascu\'es}, \citenamefont {Wang},\ and\ \citenamefont
  {Fan}}]{PRL-2022-Li}%
  \BibitemOpen
  \bibfield  {author} {\bibinfo {author} {\bibfnamefont {Z.-D.}\ \bibnamefont
  {Li}}, \bibinfo {author} {\bibfnamefont {Y.-L.}\ \bibnamefont {Mao}},
  \bibinfo {author} {\bibfnamefont {M.}~\bibnamefont {Weilenmann}}, \bibinfo
  {author} {\bibfnamefont {A.}~\bibnamefont {Tavakoli}}, \bibinfo {author}
  {\bibfnamefont {H.}~\bibnamefont {Chen}}, \bibinfo {author} {\bibfnamefont
  {L.}~\bibnamefont {Feng}}, \bibinfo {author} {\bibfnamefont {S.-J.}\
  \bibnamefont {Yang}}, \bibinfo {author} {\bibfnamefont {M.-O.}\ \bibnamefont
  {Renou}}, \bibinfo {author} {\bibfnamefont {D.}~\bibnamefont {Trillo}},
  \bibinfo {author} {\bibfnamefont {T.~P.}\ \bibnamefont {Le}}, \bibinfo
  {author} {\bibfnamefont {N.}~\bibnamefont {Gisin}}, \bibinfo {author}
  {\bibfnamefont {A.}~\bibnamefont {Ac\'{\i}n}}, \bibinfo {author}
  {\bibfnamefont {M.}~\bibnamefont {Navascu\'es}}, \bibinfo {author}
  {\bibfnamefont {Z.}~\bibnamefont {Wang}}, \ and\ \bibinfo {author}
  {\bibfnamefont {J.}~\bibnamefont {Fan}},\ }\bibinfo {title} {Testing real
  quantum theory in an optical quantum network},\ \href {\doibase
  10.1103/PhysRevLett.128.040402} {\bibfield  {journal} {\bibinfo  {journal}
  {Phys. Rev. Lett.}\ }\textbf {\bibinfo {volume} {128}},\ \bibinfo {pages}
  {040402} (\bibinfo {year} {2022}{\natexlab{a}})}\BibitemShut {NoStop}%
\bibitem [{\citenamefont {Li}\ \emph {et~al.}(2022{\natexlab{b}})\citenamefont
  {Li}, \citenamefont {Luo},\ and\ \citenamefont {Sun}}]{PRA-2022-Luo}%
  \BibitemOpen
  \bibfield  {author} {\bibinfo {author} {\bibfnamefont {N.}~\bibnamefont
  {Li}}, \bibinfo {author} {\bibfnamefont {S.}~\bibnamefont {Luo}}, \ and\
  \bibinfo {author} {\bibfnamefont {Y.}~\bibnamefont {Sun}},\ }\bibinfo {title}
  {Brukner-zeilinger invariant information in the presence of conjugate
  symmetry},\ \href {\doibase 10.1103/PhysRevA.106.032404} {\bibfield
  {journal} {\bibinfo  {journal} {Phys. Rev. A}\ }\textbf {\bibinfo {volume}
  {106}},\ \bibinfo {pages} {032404} (\bibinfo {year}
  {2022}{\natexlab{b}})}\BibitemShut {NoStop}%
\bibitem [{\citenamefont {Wu}\ \emph {et~al.}(2023)\citenamefont {Wu},
  \citenamefont {Kondra}, \citenamefont {Scandolo}, \citenamefont {Rana},
  \citenamefont {Xiang}, \citenamefont {Li}, \citenamefont {Guo},\ and\
  \citenamefont {Streltsov}}]{arXiv-2023-Guo}%
  \BibitemOpen
  \bibfield  {author} {\bibinfo {author} {\bibfnamefont {K.-D.}\ \bibnamefont
  {Wu}}, \bibinfo {author} {\bibfnamefont {T.~V.}\ \bibnamefont {Kondra}},
  \bibinfo {author} {\bibfnamefont {C.~M.}\ \bibnamefont {Scandolo}}, \bibinfo
  {author} {\bibfnamefont {S.}~\bibnamefont {Rana}}, \bibinfo {author}
  {\bibfnamefont {G.-Y.}\ \bibnamefont {Xiang}}, \bibinfo {author}
  {\bibfnamefont {C.-F.}\ \bibnamefont {Li}}, \bibinfo {author} {\bibfnamefont
  {G.-C.}\ \bibnamefont {Guo}}, \ and\ \bibinfo {author} {\bibfnamefont
  {A.}~\bibnamefont {Streltsov}},\ }\bibinfo {title} {Resource theory of
  imaginarity: New distributed scenarios},\ \href
  {https://arxiv.org/abs/2301.04782} {\bibfield  {journal} {\bibinfo  {journal}
  {arXiv preprint arXiv:2301.04782}\ } (\bibinfo {year} {2023})}\BibitemShut
  {NoStop}%
\bibitem [{\citenamefont {Nielsen}\ and\ \citenamefont
  {Chuang}(2010)}]{Nielsen-2010-quantum}%
  \BibitemOpen
  \bibfield  {author} {\bibinfo {author} {\bibfnamefont {M.~A.}\ \bibnamefont
  {Nielsen}}\ and\ \bibinfo {author} {\bibfnamefont {I.~L.}\ \bibnamefont
  {Chuang}},\ }\href@noop {} {\emph {\bibinfo {title} {Quantum computation and
  quantum information}}}\ (\bibinfo  {publisher} {Cambridge university press},\
  \bibinfo {year} {2010})\BibitemShut {NoStop}%
\bibitem [{\citenamefont {Horodecki}\ and\ \citenamefont
  {Oppenheim}(2013)}]{IJMPB-Horodechi-2013}%
  \BibitemOpen
  \bibfield  {author} {\bibinfo {author} {\bibfnamefont {M.}~\bibnamefont
  {Horodecki}}\ and\ \bibinfo {author} {\bibfnamefont {J.}~\bibnamefont
  {Oppenheim}},\ }\bibinfo {title} {(quantumness in the context of) resource
  theories},\ \href {\doibase 10.1142/S0217979213450197} {\bibfield  {journal}
  {\bibinfo  {journal} {International Journal of Modern Physics B}\ }\textbf
  {\bibinfo {volume} {27}},\ \bibinfo {pages} {1345019} (\bibinfo {year}
  {2013})}\BibitemShut {NoStop}%
\bibitem [{\citenamefont {Chitambar}\ and\ \citenamefont
  {Gour}(2019)}]{RMP-Gour-2019}%
  \BibitemOpen
  \bibfield  {author} {\bibinfo {author} {\bibfnamefont {E.}~\bibnamefont
  {Chitambar}}\ and\ \bibinfo {author} {\bibfnamefont {G.}~\bibnamefont
  {Gour}},\ }\bibinfo {title} {Quantum resource theories},\ \href {\doibase
  10.1103/RevModPhys.91.025001} {\bibfield  {journal} {\bibinfo  {journal}
  {Rev. Mod. Phys.}\ }\textbf {\bibinfo {volume} {91}},\ \bibinfo {pages}
  {025001} (\bibinfo {year} {2019})}\BibitemShut {NoStop}%
\bibitem [{\citenamefont {Vedral}\ \emph {et~al.}(1997)\citenamefont {Vedral},
  \citenamefont {Plenio}, \citenamefont {Rippin},\ and\ \citenamefont
  {Knight}}]{PRL-1997-Vedral}%
  \BibitemOpen
  \bibfield  {author} {\bibinfo {author} {\bibfnamefont {V.}~\bibnamefont
  {Vedral}}, \bibinfo {author} {\bibfnamefont {M.~B.}\ \bibnamefont {Plenio}},
  \bibinfo {author} {\bibfnamefont {M.~A.}\ \bibnamefont {Rippin}}, \ and\
  \bibinfo {author} {\bibfnamefont {P.~L.}\ \bibnamefont {Knight}},\ }\bibinfo
  {title} {Quantifying entanglement},\ \href {\doibase
  10.1103/PhysRevLett.78.2275} {\bibfield  {journal} {\bibinfo  {journal}
  {Phys. Rev. Lett.}\ }\textbf {\bibinfo {volume} {78}},\ \bibinfo {pages}
  {2275} (\bibinfo {year} {1997})}\BibitemShut {NoStop}%
\bibitem [{\citenamefont {Horodecki}\ \emph {et~al.}(2009)\citenamefont
  {Horodecki}, \citenamefont {Horodecki}, \citenamefont {Horodecki},\ and\
  \citenamefont {Horodecki}}]{RMP-2009-Horodecki}%
  \BibitemOpen
  \bibfield  {author} {\bibinfo {author} {\bibfnamefont {R.}~\bibnamefont
  {Horodecki}}, \bibinfo {author} {\bibfnamefont {P.}~\bibnamefont
  {Horodecki}}, \bibinfo {author} {\bibfnamefont {M.}~\bibnamefont
  {Horodecki}}, \ and\ \bibinfo {author} {\bibfnamefont {K.}~\bibnamefont
  {Horodecki}},\ }\bibinfo {title} {Quantum entanglement},\ \href {\doibase
  10.1103/RevModPhys.81.865} {\bibfield  {journal} {\bibinfo  {journal} {Rev.
  Mod. Phys.}\ }\textbf {\bibinfo {volume} {81}},\ \bibinfo {pages} {865}
  (\bibinfo {year} {2009})}\BibitemShut {NoStop}%
\bibitem [{\citenamefont {Baumgratz}\ \emph {et~al.}(2014)\citenamefont
  {Baumgratz}, \citenamefont {Cramer},\ and\ \citenamefont
  {Plenio}}]{PRL-Plenio-2014}%
  \BibitemOpen
  \bibfield  {author} {\bibinfo {author} {\bibfnamefont {T.}~\bibnamefont
  {Baumgratz}}, \bibinfo {author} {\bibfnamefont {M.}~\bibnamefont {Cramer}}, \
  and\ \bibinfo {author} {\bibfnamefont {M.~B.}\ \bibnamefont {Plenio}},\
  }\bibinfo {title} {Quantifying coherence},\ \href {\doibase
  10.1103/PhysRevLett.113.140401} {\bibfield  {journal} {\bibinfo  {journal}
  {Phys. Rev. Lett.}\ }\textbf {\bibinfo {volume} {113}},\ \bibinfo {pages}
  {140401} (\bibinfo {year} {2014})}\BibitemShut {NoStop}%
\bibitem [{\citenamefont {Streltsov}\ \emph {et~al.}(2017)\citenamefont
  {Streltsov}, \citenamefont {Adesso},\ and\ \citenamefont
  {Plenio}}]{RMP-Plenio-2017}%
  \BibitemOpen
  \bibfield  {author} {\bibinfo {author} {\bibfnamefont {A.}~\bibnamefont
  {Streltsov}}, \bibinfo {author} {\bibfnamefont {G.}~\bibnamefont {Adesso}}, \
  and\ \bibinfo {author} {\bibfnamefont {M.~B.}\ \bibnamefont {Plenio}},\
  }\bibinfo {title} {Colloquium: Quantum coherence as a resource},\ \href
  {\doibase 10.1103/RevModPhys.89.041003} {\bibfield  {journal} {\bibinfo
  {journal} {Rev. Mod. Phys.}\ }\textbf {\bibinfo {volume} {89}},\ \bibinfo
  {pages} {041003} (\bibinfo {year} {2017})}\BibitemShut {NoStop}%
\bibitem [{\citenamefont {Bischof}\ \emph {et~al.}(2019)\citenamefont
  {Bischof}, \citenamefont {Kampermann},\ and\ \citenamefont
  {Bru{\ss}}}]{PRL-2019-Brub}%
  \BibitemOpen
  \bibfield  {author} {\bibinfo {author} {\bibfnamefont {F.}~\bibnamefont
  {Bischof}}, \bibinfo {author} {\bibfnamefont {H.}~\bibnamefont {Kampermann}},
  \ and\ \bibinfo {author} {\bibfnamefont {D.}~\bibnamefont {Bru{\ss}}},\
  }\bibinfo {title} {Resource theory of coherence based on
  positive-operator-valued measures},\ \href {\doibase
  10.1103/PhysRevLett.123.110402} {\bibfield  {journal} {\bibinfo  {journal}
  {Phys. Rev. Lett.}\ }\textbf {\bibinfo {volume} {123}},\ \bibinfo {pages}
  {110402} (\bibinfo {year} {2019})}\BibitemShut {NoStop}%
\bibitem [{\citenamefont {Wu}\ \emph {et~al.}(2021{\natexlab{c}})\citenamefont
  {Wu}, \citenamefont {Streltsov}, \citenamefont {Regula}, \citenamefont
  {Xiang}, \citenamefont {Li},\ and\ \citenamefont {Guo}}]{AQT-2021-Wu}%
  \BibitemOpen
  \bibfield  {author} {\bibinfo {author} {\bibfnamefont {K.-D.}\ \bibnamefont
  {Wu}}, \bibinfo {author} {\bibfnamefont {A.}~\bibnamefont {Streltsov}},
  \bibinfo {author} {\bibfnamefont {B.}~\bibnamefont {Regula}}, \bibinfo
  {author} {\bibfnamefont {G.-Y.}\ \bibnamefont {Xiang}}, \bibinfo {author}
  {\bibfnamefont {C.-F.}\ \bibnamefont {Li}}, \ and\ \bibinfo {author}
  {\bibfnamefont {G.-C.}\ \bibnamefont {Guo}},\ }\bibinfo {title} {Experimental
  progress on quantum coherence: Detection, quantification, and manipulation},\
  \href {\doibase https://doi.org/10.1002/qute.202100040} {\bibfield  {journal}
  {\bibinfo  {journal} {Advanced Quantum Technologies}\ }\textbf {\bibinfo
  {volume} {4}},\ \bibinfo {pages} {2100040} (\bibinfo {year}
  {2021}{\natexlab{c}})}\BibitemShut {NoStop}%
\bibitem [{\citenamefont {Goold}\ \emph {et~al.}(2016)\citenamefont {Goold},
  \citenamefont {Huber}, \citenamefont {Riera}, \citenamefont {del Rio},\ and\
  \citenamefont {Skrzypczyk}}]{Goold-2016}%
  \BibitemOpen
  \bibfield  {author} {\bibinfo {author} {\bibfnamefont {J.}~\bibnamefont
  {Goold}}, \bibinfo {author} {\bibfnamefont {M.}~\bibnamefont {Huber}},
  \bibinfo {author} {\bibfnamefont {A.}~\bibnamefont {Riera}}, \bibinfo
  {author} {\bibfnamefont {L.}~\bibnamefont {del Rio}}, \ and\ \bibinfo
  {author} {\bibfnamefont {P.}~\bibnamefont {Skrzypczyk}},\ }\bibinfo {title}
  {The role of quantum information in thermodynamics-a topical review},\ \href
  {\doibase 10.1088/1751-8113/49/14/143001} {\bibfield  {journal} {\bibinfo
  {journal} {Journal of Physics A: Mathematical and Theoretical}\ }\textbf
  {\bibinfo {volume} {49}},\ \bibinfo {pages} {143001} (\bibinfo {year}
  {2016})}\BibitemShut {NoStop}%
\bibitem [{\citenamefont {Lostaglio}(2019)}]{lostaglio2019introductory}%
  \BibitemOpen
  \bibfield  {author} {\bibinfo {author} {\bibfnamefont {M.}~\bibnamefont
  {Lostaglio}},\ }\bibinfo {title} {An introductory review of the resource
  theory approach to thermodynamics},\ \href
  {https://iopscience.iop.org/article/10.1088/1361-6633/ab46e5/meta} {\bibfield
   {journal} {\bibinfo  {journal} {Reports on Progress in Physics}\ }\textbf
  {\bibinfo {volume} {82}},\ \bibinfo {pages} {114001} (\bibinfo {year}
  {2019})}\BibitemShut {NoStop}%
\bibitem [{\citenamefont {Horodecki}\ \emph {et~al.}(2003)\citenamefont
  {Horodecki}, \citenamefont {Horodecki},\ and\ \citenamefont
  {Oppenheim}}]{PRA-2003-Horodecki}%
  \BibitemOpen
  \bibfield  {author} {\bibinfo {author} {\bibfnamefont {M.}~\bibnamefont
  {Horodecki}}, \bibinfo {author} {\bibfnamefont {P.}~\bibnamefont
  {Horodecki}}, \ and\ \bibinfo {author} {\bibfnamefont {J.}~\bibnamefont
  {Oppenheim}},\ }\bibinfo {title} {Reversible transformations from pure to
  mixed states and the unique measure of information},\ \href {\doibase
  10.1103/PhysRevA.67.062104} {\bibfield  {journal} {\bibinfo  {journal} {Phys.
  Rev. A}\ }\textbf {\bibinfo {volume} {67}},\ \bibinfo {pages} {062104}
  (\bibinfo {year} {2003})}\BibitemShut {NoStop}%
\bibitem [{\citenamefont {Gour}\ \emph {et~al.}(2015)\citenamefont {Gour},
  \citenamefont {M\"{u}ller}, \citenamefont {Narasimhachar}, \citenamefont
  {Spekkens},\ and\ \citenamefont {Halpern}}]{PR-2015-Gour}%
  \BibitemOpen
  \bibfield  {author} {\bibinfo {author} {\bibfnamefont {G.}~\bibnamefont
  {Gour}}, \bibinfo {author} {\bibfnamefont {M.~P.}\ \bibnamefont
  {M\"{u}ller}}, \bibinfo {author} {\bibfnamefont {V.}~\bibnamefont
  {Narasimhachar}}, \bibinfo {author} {\bibfnamefont {R.~W.}\ \bibnamefont
  {Spekkens}}, \ and\ \bibinfo {author} {\bibfnamefont {N.~Y.}\ \bibnamefont
  {Halpern}},\ }\bibinfo {title} {The resource theory of informational
  nonequilibrium in thermodynamics},\ \href {\doibase
  https://doi.org/10.1016/j.physrep.2015.04.003} {\bibfield  {journal}
  {\bibinfo  {journal} {Physics Reports}\ }\textbf {\bibinfo {volume} {583}},\
  \bibinfo {pages} {1} (\bibinfo {year} {2015})},\ \bibinfo {note} {the
  resource theory of informational nonequilibrium in
  thermodynamics}\BibitemShut {NoStop}%
\bibitem [{\citenamefont {Streltsov}\ \emph {et~al.}(2018)\citenamefont
  {Streltsov}, \citenamefont {Kampermann}, \citenamefont {W\"{o}lk},
  \citenamefont {Gessner},\ and\ \citenamefont
  {Bru{\ss}}}]{NJP-2018-Streltsov}%
  \BibitemOpen
  \bibfield  {author} {\bibinfo {author} {\bibfnamefont {A.}~\bibnamefont
  {Streltsov}}, \bibinfo {author} {\bibfnamefont {H.}~\bibnamefont
  {Kampermann}}, \bibinfo {author} {\bibfnamefont {S.}~\bibnamefont
  {W\"{o}lk}}, \bibinfo {author} {\bibfnamefont {M.}~\bibnamefont {Gessner}}, \
  and\ \bibinfo {author} {\bibfnamefont {D.}~\bibnamefont {Bru{\ss}}},\
  }\bibinfo {title} {Maximal coherence and the resource theory of purity},\
  \href {\doibase 10.1088/1367-2630/aac484} {\bibfield  {journal} {\bibinfo
  {journal} {New Journal of Physics}\ }\textbf {\bibinfo {volume} {20}},\
  \bibinfo {pages} {053058} (\bibinfo {year} {2018})}\BibitemShut {NoStop}%
\bibitem [{\citenamefont {De~Vicente}(2014)}]{JPA-2014-De}%
  \BibitemOpen
  \bibfield  {author} {\bibinfo {author} {\bibfnamefont {J.~I.}\ \bibnamefont
  {De~Vicente}},\ }\bibinfo {title} {On nonlocality as a resource theory and
  nonlocality measures},\ \href
  {https://iopscience.iop.org/article/10.1088/1751-8113/47/42/424017/meta}
  {\bibfield  {journal} {\bibinfo  {journal} {Journal of Physics A:
  Mathematical and Theoretical}\ }\textbf {\bibinfo {volume} {47}},\ \bibinfo
  {pages} {424017} (\bibinfo {year} {2014})}\BibitemShut {NoStop}%
\bibitem [{\citenamefont {Gianfelici}\ \emph {et~al.}(2021)\citenamefont
  {Gianfelici}, \citenamefont {Kampermann},\ and\ \citenamefont
  {Bru{\ss}}}]{NJP-2021-Brub}%
  \BibitemOpen
  \bibfield  {author} {\bibinfo {author} {\bibfnamefont {G.}~\bibnamefont
  {Gianfelici}}, \bibinfo {author} {\bibfnamefont {H.}~\bibnamefont
  {Kampermann}}, \ and\ \bibinfo {author} {\bibfnamefont {D.}~\bibnamefont
  {Bru{\ss}}},\ }\bibinfo {title} {Hierarchy of continuous-variable quantum
  resource theories},\ \href
  {https://iopscience.iop.org/article/10.1088/1367-2630/ac2f90/meta} {\bibfield
   {journal} {\bibinfo  {journal} {New Journal of Physics}\ }\textbf {\bibinfo
  {volume} {23}},\ \bibinfo {pages} {113008} (\bibinfo {year}
  {2021})}\BibitemShut {NoStop}%
\bibitem [{\citenamefont {Regula}\ \emph {et~al.}(2021)\citenamefont {Regula},
  \citenamefont {Lami}, \citenamefont {Ferrari},\ and\ \citenamefont
  {Takagi}}]{PRL-Takagi-2021}%
  \BibitemOpen
  \bibfield  {author} {\bibinfo {author} {\bibfnamefont {B.}~\bibnamefont
  {Regula}}, \bibinfo {author} {\bibfnamefont {L.}~\bibnamefont {Lami}},
  \bibinfo {author} {\bibfnamefont {G.}~\bibnamefont {Ferrari}}, \ and\
  \bibinfo {author} {\bibfnamefont {R.}~\bibnamefont {Takagi}},\ }\bibinfo
  {title} {Operational quantification of continuous-variable quantum
  resources},\ \href {\doibase 10.1103/PhysRevLett.126.110403} {\bibfield
  {journal} {\bibinfo  {journal} {Phys. Rev. Lett.}\ }\textbf {\bibinfo
  {volume} {126}},\ \bibinfo {pages} {110403} (\bibinfo {year}
  {2021})}\BibitemShut {NoStop}%
\bibitem [{\citenamefont {Zhang}\ \emph {et~al.}(2016)\citenamefont {Zhang},
  \citenamefont {Shao}, \citenamefont {Li},\ and\ \citenamefont
  {Fan}}]{PRA-Fan-2016}%
  \BibitemOpen
  \bibfield  {author} {\bibinfo {author} {\bibfnamefont {Y.-R.}\ \bibnamefont
  {Zhang}}, \bibinfo {author} {\bibfnamefont {L.-H.}\ \bibnamefont {Shao}},
  \bibinfo {author} {\bibfnamefont {Y.}~\bibnamefont {Li}}, \ and\ \bibinfo
  {author} {\bibfnamefont {H.}~\bibnamefont {Fan}},\ }\bibinfo {title}
  {Quantifying coherence in infinite-dimensional systems},\ \href {\doibase
  10.1103/PhysRevA.93.012334} {\bibfield  {journal} {\bibinfo  {journal} {Phys.
  Rev. A}\ }\textbf {\bibinfo {volume} {93}},\ \bibinfo {pages} {012334}
  (\bibinfo {year} {2016})}\BibitemShut {NoStop}%
\bibitem [{\citenamefont {Xu}(2016)}]{PRA-Xu-2016}%
  \BibitemOpen
  \bibfield  {author} {\bibinfo {author} {\bibfnamefont {J.}~\bibnamefont
  {Xu}},\ }\bibinfo {title} {Quantifying coherence of gaussian states},\ \href
  {\doibase 10.1103/PhysRevA.93.032111} {\bibfield  {journal} {\bibinfo
  {journal} {Phys. Rev. A}\ }\textbf {\bibinfo {volume} {93}},\ \bibinfo
  {pages} {032111} (\bibinfo {year} {2016})}\BibitemShut {NoStop}%
\bibitem [{\citenamefont {Albarelli}\ \emph {et~al.}(2017)\citenamefont
  {Albarelli}, \citenamefont {Genoni},\ and\ \citenamefont
  {Paris}}]{PRA-2017-Paris}%
  \BibitemOpen
  \bibfield  {author} {\bibinfo {author} {\bibfnamefont {F.}~\bibnamefont
  {Albarelli}}, \bibinfo {author} {\bibfnamefont {M.~G.}\ \bibnamefont
  {Genoni}}, \ and\ \bibinfo {author} {\bibfnamefont {M.~G.~A.}\ \bibnamefont
  {Paris}},\ }\bibinfo {title} {Generation of coherence via gaussian
  measurements},\ \href {\doibase 10.1103/PhysRevA.96.012337} {\bibfield
  {journal} {\bibinfo  {journal} {Phys. Rev. A}\ }\textbf {\bibinfo {volume}
  {96}},\ \bibinfo {pages} {012337} (\bibinfo {year} {2017})}\BibitemShut
  {NoStop}%
\bibitem [{\citenamefont {Braunstein}\ and\ \citenamefont {van
  Loock}(2005)}]{RMP-2005-Braunstein}%
  \BibitemOpen
  \bibfield  {author} {\bibinfo {author} {\bibfnamefont {S.~L.}\ \bibnamefont
  {Braunstein}}\ and\ \bibinfo {author} {\bibfnamefont {P.}~\bibnamefont {van
  Loock}},\ }\bibinfo {title} {Quantum information with continuous variables},\
  \href {\doibase 10.1103/RevModPhys.77.513} {\bibfield  {journal} {\bibinfo
  {journal} {Rev. Mod. Phys.}\ }\textbf {\bibinfo {volume} {77}},\ \bibinfo
  {pages} {513} (\bibinfo {year} {2005})}\BibitemShut {NoStop}%
\bibitem [{\citenamefont {Wang}\ \emph {et~al.}(2007)\citenamefont {Wang},
  \citenamefont {Hiroshima}, \citenamefont {Tomita},\ and\ \citenamefont
  {Hayashi}}]{PR-2007-Wang}%
  \BibitemOpen
  \bibfield  {author} {\bibinfo {author} {\bibfnamefont {X.-B.}\ \bibnamefont
  {Wang}}, \bibinfo {author} {\bibfnamefont {T.}~\bibnamefont {Hiroshima}},
  \bibinfo {author} {\bibfnamefont {A.}~\bibnamefont {Tomita}}, \ and\ \bibinfo
  {author} {\bibfnamefont {M.}~\bibnamefont {Hayashi}},\ }\bibinfo {title}
  {Quantum information with gaussian states},\ \href {\doibase
  https://doi.org/10.1016/j.physrep.2007.04.005} {\bibfield  {journal}
  {\bibinfo  {journal} {Physics Reports}\ }\textbf {\bibinfo {volume} {448}},\
  \bibinfo {pages} {1} (\bibinfo {year} {2007})}\BibitemShut {NoStop}%
\bibitem [{\citenamefont {Ferraro}\ \emph {et~al.}(2005)\citenamefont
  {Ferraro}, \citenamefont {Olivares},\ and\ \citenamefont
  {Paris}}]{arXiv-2005-Ferraro}%
  \BibitemOpen
  \bibfield  {author} {\bibinfo {author} {\bibfnamefont {A.}~\bibnamefont
  {Ferraro}}, \bibinfo {author} {\bibfnamefont {S.}~\bibnamefont {Olivares}}, \
  and\ \bibinfo {author} {\bibfnamefont {M.~G.}\ \bibnamefont {Paris}},\
  }\bibinfo {title} {Gaussian states in continuous variable quantum
  information},\ \href {https://arxiv.org/abs/quant-ph/0503237} {\bibfield
  {journal} {\bibinfo  {journal} {arXiv preprint quant-ph/0503237}\ } (\bibinfo
  {year} {2005})}\BibitemShut {NoStop}%
\bibitem [{\citenamefont {Olivares}(2012)}]{EPJ-2012-Olivares}%
  \BibitemOpen
  \bibfield  {author} {\bibinfo {author} {\bibfnamefont {S.}~\bibnamefont
  {Olivares}},\ }\bibinfo {title} {Quantum optics in the phase space: a
  tutorial on gaussian states},\ \href
  {https://link.springer.com/article/10.1140/epjst/e2012-01532-4} {\bibfield
  {journal} {\bibinfo  {journal} {The European Physical Journal Special
  Topics}\ }\textbf {\bibinfo {volume} {203}},\ \bibinfo {pages} {3} (\bibinfo
  {year} {2012})}\BibitemShut {NoStop}%
\bibitem [{\citenamefont {Weedbrook}\ \emph {et~al.}(2012)\citenamefont
  {Weedbrook}, \citenamefont {Pirandola}, \citenamefont {Garc\'{\i}a-Patr\'on},
  \citenamefont {Cerf}, \citenamefont {Ralph}, \citenamefont {Shapiro},\ and\
  \citenamefont {Lloyd}}]{RMP-2012-Weedbrook}%
  \BibitemOpen
  \bibfield  {author} {\bibinfo {author} {\bibfnamefont {C.}~\bibnamefont
  {Weedbrook}}, \bibinfo {author} {\bibfnamefont {S.}~\bibnamefont
  {Pirandola}}, \bibinfo {author} {\bibfnamefont {R.}~\bibnamefont
  {Garc\'{\i}a-Patr\'on}}, \bibinfo {author} {\bibfnamefont {N.~J.}\
  \bibnamefont {Cerf}}, \bibinfo {author} {\bibfnamefont {T.~C.}\ \bibnamefont
  {Ralph}}, \bibinfo {author} {\bibfnamefont {J.~H.}\ \bibnamefont {Shapiro}},
  \ and\ \bibinfo {author} {\bibfnamefont {S.}~\bibnamefont {Lloyd}},\
  }\bibinfo {title} {Gaussian quantum information},\ \href {\doibase
  10.1103/RevModPhys.84.621} {\bibfield  {journal} {\bibinfo  {journal} {Rev.
  Mod. Phys.}\ }\textbf {\bibinfo {volume} {84}},\ \bibinfo {pages} {621}
  (\bibinfo {year} {2012})}\BibitemShut {NoStop}%
\bibitem [{\citenamefont {Adesso}\ \emph {et~al.}(2014)\citenamefont {Adesso},
  \citenamefont {Ragy},\ and\ \citenamefont {Lee}}]{OSID-2014-Adesso}%
  \BibitemOpen
  \bibfield  {author} {\bibinfo {author} {\bibfnamefont {G.}~\bibnamefont
  {Adesso}}, \bibinfo {author} {\bibfnamefont {S.}~\bibnamefont {Ragy}}, \ and\
  \bibinfo {author} {\bibfnamefont {A.~R.}\ \bibnamefont {Lee}},\ }\bibinfo
  {title} {Continuous variable quantum information: Gaussian states and
  beyond},\ \href {\doibase 10.1142/S1230161214400010} {\bibfield  {journal}
  {\bibinfo  {journal} {Open Systems \& Information Dynamics}\ }\textbf
  {\bibinfo {volume} {21}},\ \bibinfo {pages} {1440001} (\bibinfo {year}
  {2014})}\BibitemShut {NoStop}%
\bibitem [{\citenamefont {Serafini}(2017)}]{book-2017-Serafini}%
  \BibitemOpen
  \bibfield  {author} {\bibinfo {author} {\bibfnamefont {A.}~\bibnamefont
  {Serafini}},\ }\href@noop {} {\emph {\bibinfo {title} {Quantum continuous
  variables: a primer of theoretical methods}}}\ (\bibinfo  {publisher} {CRC
  press},\ \bibinfo {year} {2017})\BibitemShut {NoStop}%
\bibitem [{\citenamefont {Uhlmann}(1976)}]{RMP-1976-Uhlmann}%
  \BibitemOpen
  \bibfield  {author} {\bibinfo {author} {\bibfnamefont {A.}~\bibnamefont
  {Uhlmann}},\ }\bibinfo {title} {The ¡°transition probability¡± in the state
  space of a $\ast$-algebra},\ \href
  {https://www.sciencedirect.com/science/article/abs/pii/0034487776900604}
  {\bibfield  {journal} {\bibinfo  {journal} {Reports on Mathematical Physics}\
  }\textbf {\bibinfo {volume} {9}},\ \bibinfo {pages} {273} (\bibinfo {year}
  {1976})}\BibitemShut {NoStop}%
\bibitem [{\citenamefont {Jozsa}(1994)}]{JMP-1994-Jozsa}%
  \BibitemOpen
  \bibfield  {author} {\bibinfo {author} {\bibfnamefont {R.}~\bibnamefont
  {Jozsa}},\ }\bibinfo {title} {Fidelity for mixed quantum states},\ \href
  {\doibase 10.1080/09500349414552171} {\bibfield  {journal} {\bibinfo
  {journal} {Journal of Modern Optics}\ }\textbf {\bibinfo {volume} {41}},\
  \bibinfo {pages} {2315} (\bibinfo {year} {1994})}\BibitemShut {NoStop}%
\bibitem [{\citenamefont {Quesada}\ \emph {et~al.}(2019)\citenamefont
  {Quesada}, \citenamefont {Helt}, \citenamefont {Izaac}, \citenamefont
  {Arrazola}, \citenamefont {Shahrokhshahi}, \citenamefont {Myers},\ and\
  \citenamefont {Sabapathy}}]{PRA-2019-Quesada}%
  \BibitemOpen
  \bibfield  {author} {\bibinfo {author} {\bibfnamefont {N.}~\bibnamefont
  {Quesada}}, \bibinfo {author} {\bibfnamefont {L.~G.}\ \bibnamefont {Helt}},
  \bibinfo {author} {\bibfnamefont {J.}~\bibnamefont {Izaac}}, \bibinfo
  {author} {\bibfnamefont {J.~M.}\ \bibnamefont {Arrazola}}, \bibinfo {author}
  {\bibfnamefont {R.}~\bibnamefont {Shahrokhshahi}}, \bibinfo {author}
  {\bibfnamefont {C.~R.}\ \bibnamefont {Myers}}, \ and\ \bibinfo {author}
  {\bibfnamefont {K.~K.}\ \bibnamefont {Sabapathy}},\ }\bibinfo {title}
  {Simulating realistic non-gaussian state preparation},\ \href {\doibase
  10.1103/PhysRevA.100.022341} {\bibfield  {journal} {\bibinfo  {journal}
  {Phys. Rev. A}\ }\textbf {\bibinfo {volume} {100}},\ \bibinfo {pages}
  {022341} (\bibinfo {year} {2019})}\BibitemShut {NoStop}%
\bibitem [{\citenamefont {Banchi}\ \emph {et~al.}(2015)\citenamefont {Banchi},
  \citenamefont {Braunstein},\ and\ \citenamefont
  {Pirandola}}]{PRL-2015-Banchi}%
  \BibitemOpen
  \bibfield  {author} {\bibinfo {author} {\bibfnamefont {L.}~\bibnamefont
  {Banchi}}, \bibinfo {author} {\bibfnamefont {S.~L.}\ \bibnamefont
  {Braunstein}}, \ and\ \bibinfo {author} {\bibfnamefont {S.}~\bibnamefont
  {Pirandola}},\ }\bibinfo {title} {Quantum fidelity for arbitrary gaussian
  states},\ \href {\doibase 10.1103/PhysRevLett.115.260501} {\bibfield
  {journal} {\bibinfo  {journal} {Phys. Rev. Lett.}\ }\textbf {\bibinfo
  {volume} {115}},\ \bibinfo {pages} {260501} (\bibinfo {year}
  {2015})}\BibitemShut {NoStop}%
\bibitem [{\citenamefont {Simon}\ \emph {et~al.}(1994)\citenamefont {Simon},
  \citenamefont {Mukunda},\ and\ \citenamefont {Dutta}}]{PRA-1994-Simon}%
  \BibitemOpen
  \bibfield  {author} {\bibinfo {author} {\bibfnamefont {R.}~\bibnamefont
  {Simon}}, \bibinfo {author} {\bibfnamefont {N.}~\bibnamefont {Mukunda}}, \
  and\ \bibinfo {author} {\bibfnamefont {B.}~\bibnamefont {Dutta}},\ }\bibinfo
  {title} {Quantum-noise matrix for multimode systems: U(n) invariance,
  squeezing, and normal forms},\ \href {\doibase 10.1103/PhysRevA.49.1567}
  {\bibfield  {journal} {\bibinfo  {journal} {Phys. Rev. A}\ }\textbf {\bibinfo
  {volume} {49}},\ \bibinfo {pages} {1567} (\bibinfo {year}
  {1994})}\BibitemShut {NoStop}%
\bibitem [{\citenamefont {Scutaru}(1998)}]{JPA-1998-Scutaru}%
  \BibitemOpen
  \bibfield  {author} {\bibinfo {author} {\bibfnamefont {H.}~\bibnamefont
  {Scutaru}},\ }\bibinfo {title} {Fidelity for displaced squeezed thermal
  states and the oscillator semigroup},\ \href {\doibase
  10.1088/0305-4470/31/15/025} {\bibfield  {journal} {\bibinfo  {journal}
  {Journal of Physics A: Mathematical and General}\ }\textbf {\bibinfo {volume}
  {31}},\ \bibinfo {pages} {3659} (\bibinfo {year} {1998})}\BibitemShut
  {NoStop}%
\bibitem [{\citenamefont {Marian}\ \emph {et~al.}(2003)\citenamefont {Marian},
  \citenamefont {Marian},\ and\ \citenamefont {Scutaru}}]{PRA-2003-Marian}%
  \BibitemOpen
  \bibfield  {author} {\bibinfo {author} {\bibfnamefont {P.}~\bibnamefont
  {Marian}}, \bibinfo {author} {\bibfnamefont {T.~A.}\ \bibnamefont {Marian}},
  \ and\ \bibinfo {author} {\bibfnamefont {H.}~\bibnamefont {Scutaru}},\
  }\bibinfo {title} {Bures distance as a measure of entanglement for two-mode
  squeezed thermal states},\ \href {\doibase 10.1103/PhysRevA.68.062309}
  {\bibfield  {journal} {\bibinfo  {journal} {Phys. Rev. A}\ }\textbf {\bibinfo
  {volume} {68}},\ \bibinfo {pages} {062309} (\bibinfo {year}
  {2003})}\BibitemShut {NoStop}%
\bibitem [{\citenamefont {Marian}\ and\ \citenamefont
  {Marian}(2008)}]{PRA-2008-Marian}%
  \BibitemOpen
  \bibfield  {author} {\bibinfo {author} {\bibfnamefont {P.}~\bibnamefont
  {Marian}}\ and\ \bibinfo {author} {\bibfnamefont {T.~A.}\ \bibnamefont
  {Marian}},\ }\bibinfo {title} {Bures distance as a measure of entanglement
  for symmetric two-mode gaussian states},\ \href {\doibase
  10.1103/PhysRevA.77.062319} {\bibfield  {journal} {\bibinfo  {journal} {Phys.
  Rev. A}\ }\textbf {\bibinfo {volume} {77}},\ \bibinfo {pages} {062319}
  (\bibinfo {year} {2008})}\BibitemShut {NoStop}%
\bibitem [{\citenamefont {Nha}\ and\ \citenamefont
  {Carmichael}(2005)}]{PRA-2005-Nha}%
  \BibitemOpen
  \bibfield  {author} {\bibinfo {author} {\bibfnamefont {H.}~\bibnamefont
  {Nha}}\ and\ \bibinfo {author} {\bibfnamefont {H.~J.}\ \bibnamefont
  {Carmichael}},\ }\bibinfo {title} {Distinguishing two single-mode gaussian
  states by homodyne detection: An information-theoretic approach},\ \href
  {\doibase 10.1103/PhysRevA.71.032336} {\bibfield  {journal} {\bibinfo
  {journal} {Phys. Rev. A}\ }\textbf {\bibinfo {volume} {71}},\ \bibinfo
  {pages} {032336} (\bibinfo {year} {2005})}\BibitemShut {NoStop}%
\bibitem [{\citenamefont {Olivares}\ \emph {et~al.}(2006)\citenamefont
  {Olivares}, \citenamefont {Paris},\ and\ \citenamefont
  {Andersen}}]{PRA-2006-Olivares}%
  \BibitemOpen
  \bibfield  {author} {\bibinfo {author} {\bibfnamefont {S.}~\bibnamefont
  {Olivares}}, \bibinfo {author} {\bibfnamefont {M.~G.~A.}\ \bibnamefont
  {Paris}}, \ and\ \bibinfo {author} {\bibfnamefont {U.~L.}\ \bibnamefont
  {Andersen}},\ }\bibinfo {title} {Cloning of gaussian states by linear
  optics},\ \href {\doibase 10.1103/PhysRevA.73.062330} {\bibfield  {journal}
  {\bibinfo  {journal} {Phys. Rev. A}\ }\textbf {\bibinfo {volume} {73}},\
  \bibinfo {pages} {062330} (\bibinfo {year} {2006})}\BibitemShut {NoStop}%
\bibitem [{\citenamefont {Marian}\ and\ \citenamefont
  {Marian}(2012)}]{PRA-2012-Marian}%
  \BibitemOpen
  \bibfield  {author} {\bibinfo {author} {\bibfnamefont {P.}~\bibnamefont
  {Marian}}\ and\ \bibinfo {author} {\bibfnamefont {T.~A.}\ \bibnamefont
  {Marian}},\ }\bibinfo {title} {Uhlmann fidelity between two-mode gaussian
  states},\ \href {\doibase 10.1103/PhysRevA.86.022340} {\bibfield  {journal}
  {\bibinfo  {journal} {Phys. Rev. A}\ }\textbf {\bibinfo {volume} {86}},\
  \bibinfo {pages} {022340} (\bibinfo {year} {2012})}\BibitemShut {NoStop}%
\bibitem [{\citenamefont {Folland}(1989)}]{book-1989-Folland}%
  \BibitemOpen
  \bibfield  {author} {\bibinfo {author} {\bibfnamefont {G.~B.}\ \bibnamefont
  {Folland}},\ }\href@noop {} {\emph {\bibinfo {title} {Harmonic analysis in
  phase space}}},\ \bibinfo {number} {pages 256-257, Appendix A, Theorem 1}\
  (\bibinfo  {publisher} {Princeton university press},\ \bibinfo {year}
  {1989})\BibitemShut {NoStop}%
\end{thebibliography}

%

\end{document}